\documentclass[a4paper,fleqn,usenatbib]{mnras}

\usepackage{newtxtext,newtxmath}

\usepackage[T1]{fontenc}
\usepackage{ae,aecompl}

\usepackage{graphicx}	
\usepackage{amsmath}	
\usepackage{nccmath}
\usepackage{amssymb}	
\usepackage{verbatim} 
\usepackage{subcaption}
\usepackage{placeins}
\def \Msun{\ {\rm M_\odot}}

\def \lcdm{$\Lambda$CDM}
\def \lcdmdm{$\Lambda$CDM$_{\text{DM}}$}
\def \lcdmg{$\Lambda$CDM$_{\text{DM+Gas}}$}
\def \lwdm{$\Lambda$WDM}
\def \lwdmdm{$\Lambda$WDM$_{\text{DM}}$}
\def \qcdm{$\phi$CDM}
\def \qcdmdm{$\phi$CDM$_{\text{DM}}$}
\def \qcdmg{$\phi$CDM$_{\text{DM+Gas}}$}

\def \cdeodmg{CDE50$_{\text{DM+Gas}}$}

\def \cdeoodmg{CDE99$_{\text{DM+Gas}}$}
\def \e3{$\textbf{e}_{3}$}

\title[Cosmological Signatures of the Dark Sector ]{Cosmological Signatures of Dark Sector Physics: The Evolution of Haloes and Spin Alignment}
\author[A. W. Jibrail et al.]{
Absem W. Jibrail,$^{1}$\thanks{E-mail: ajib0457@uni.sydney.edu.au (AWJ)}
Pascal J. Elahi$^{2}$ and Geraint F. Lewis$^{1}$ 
\\
$^{1}$Sydney Institute for Astronomy, School of Physics, A28, The University of Sydney, NSW, 2006, Australia\\
$^{2}$International Centre for Radio Astronomy Research (ICRAR), The University of Western Australia, 35 Stirling Hwy, \\
Crawley, Western Australia 6009, Australia}

\date{Accepted XXX. Received YYY; in original form ZZZ}

\pubyear{2019}
\begin{document}
\label{firstpage}
\pagerange{\pageref{firstpage}--\pageref{lastpage}}
\maketitle
\begin{abstract}
The standard cosmological paradigm currently lacks a detailed account of physics in the dark sector, the dark matter and energy that dominate cosmic evolution.
In this paper, we consider the 
distinguishing factors between three alternative models - warm dark matter, quintessence and coupled dark matter-energy - and \lcdm{}
through numerical simulations of cosmological structure formation.
 Key halo statistics - halo spin/velocity alignment between large-scale structure and neighboring haloes, halo formation time and migration - were compared across cosmologies within the redshift range 0$\leq$$z$$\leq$2.98. We found the alignment of halo motion and spin to large-scale structures and neighbouring haloes to be similar in all cosmologies for a range of redshifts. The search was extended to low density regions, avoiding non-linear disturbances of halo spins, yet very similar alignment trends were found between cosmologies which are difficult to characterize and use as a probe of cosmology. 
 We found haloes in quintessence cosmologies form earlier than their \lcdm{} counterparts. Relating this to the fact that such haloes originate in high density regions, such findings could hold clues to distinguishing factors for the quintessence cosmology from the standard model. Although in general, halo statistics are not an accurate probe of the dark sector physics.

\end{abstract}




\section{Introduction}\label{intro}

The standard model of cosmology (\lcdm{}) is dominated by the dark sector: a Cosmological Constant, $\Lambda$, also known as Dark Energy (DE), is considered to be a negative pressure vacuum energy responsible for the late-time accelerated expansion of the Universe. Additionally, the gravitational dominance of Cold Dark Matter (CDM)  accounts for the small-scale clustering of baryonic matter. 
As the most successful cosmological model, \lcdm{} 
is well supported by large-scale observations in the anisotropies of the Cosmic Microwave Background \citep[CMB;][]{Bennett_13,Plank_14b,Plank_16}, features in the  large-scale structure of the universe \citep[LSS;][]{Abazajian_09}, Baryonic Acoustic Oscillations \citep{Beutler_11} and weak lensing \citep{Kilbinger_13}. Our understanding of galaxy clustering and structure formation has led us to rule out Hot Dark Matter (in the form of massive neutrinos) as a candidate \citep{White_83}, and CDM has well-motivated candidates from particle physics over more energetic dark matter candidates \citep{Bertone_05,Petraki_13}. Additionally, DE is well supported in the form of a cosmological constant with an equation of state $w=\rho/\text{p}\sim$ -1 \citep{Suzuki_12,Chuang_16}.

Despite its predictive success, \lcdm{} suffers from observational discrepancies and conceptual shortfalls regarding the dark sector physics. Underlying issues within \lcdm{} have motivated non-standard cosmological models, including Warm Dark Matter models (\lwdm{}), which attempts to alleviate the Missing-Satellite problem (whereby \lcdm{} produces too many satellites around central galaxies \citep{Klypin_99,Moore_99}). Although it has been proposed that this may not be an issue of \lcdm{}, but instead the result of limitations manifest within dark matter only simulations \citep{Wetzel_16}, such as their lack of feedback processes \citep[e.g][]{Bullock_00} in turn causing suppression of gas accretion for low-mass haloes. Observations also show that Milky Way satellites exhibit a dynamically stable planar distribution, known as the Vast Polar Structure \citep[e.g][]{Pawloski_12}, which is also found for Andromeda (M31) satellites \citep{Ibata_13} but are not predicted by \lcdm{} simulations nor are they accounted for in \lwdm{}. Quintessence models (\qcdm{}) alleviate the \textit{fine-tuning} problem of the initial value for vacuum energy density by substituting the cosmological constant with a scalar field \citep{Tsujikawa_13}. There are also conceptual problems with \lcdm{} such as the coincidence problem which states the energy-density of DM and DE are coincidentally similar at present day. This is unlikely given their independence, thus there may be some inter-dependence within the dark sector that alleviates the otherwise unlikely coincidence. These conceptual shortcomings motivated coupled dark sector cosmology (CDE); see \citet{Bull_16} for a comprehensive review on \lcdm{} shortfalls. 

Non-standard cosmologies do well to reproduce the predictions made by \lcdm{} whilst relieving its observational tensions aforementioned. This does not suggest the non-standard cosmologies will not vary on LSS and sub-LSS scales such as halo spin. The manner in which haloes acquire their initial spin can be described by the Tidal Torque Theory (TTT): haloes are spun-up through interactions with the tidal field, which has a significant influence on the halo before turn-around \citep{hoyle_1949,Peebles_69,Zeldovich_70}. 
It is argued that since the tidal field is the manifestation of the cosmological environment, and this field is predominantly responsible for the acquisition of initial halo spins, then the signatures of cosmology should be imprinted on the spins of galaxies \citep{Lee_pen_00}. However, denser filaments for instance, are not only governed by their underlying cosmology but by complex baryonic physics and coupling due to non-linear evolution which could mask or possibly erase any cosmological signatures in-printed on haloes therein. Haloes within voids, being less prone to such complex physics, could be useful probes however, we found that they do not constitute a large enough sample population for statistical measurements.

Observed evidence of halo spin alignments with their LSS reaffirms the utility of spin-alignment as a cosmological probe. \citet{Pen_00} shows tentative alignment in spirals. \citet{Lee_Erdogdu_07} demonstrates that galaxies in the Tully catalog are weakly orthogonal with their environment, stating an average correlation parameter ($c$) value of $c=0.084\pm 0.014$ (where $c>0$ represents an orthogonal alignment) to their LSS with 99.99$\%$ confidence that the null hypothesis of no spin-shear correlation is rejected. Galaxies observed as part of the Sloan Digital Sky Survey (SDSS) were also found to have non-random spin alignments, some of which are oriented perpendicular to their host filament \citep{Jones_10}. 

Additional simulation studies also detect significant halo spin alignments and use them to probe LSS: \citet{Faltenbacher_02,Calvo_07,Hahn_07}. \citet{Zhang_09} find clear correlation signals between LSS axes and halo spins, which can be linked to the formation of LSS and its influence on the spin and shapes of haloes. It was found that filaments are growing in width over time, where parallel spin alignment to filaments was found to be stronger at small smoothing scales and high redshift, and weaker at low redshift \citep{Trowland_13}. This demonstrates that spin alignment is sensitive to LSS evolution, which may vary within alternative cosmologies. Furthermore, spin alignment is a good tracer of merger history: \citet{Welker_14} find the more mergers a halo has undergone, the more orthogonal is its spin with respect to its host filament. \citet{Wang_17} show that haloes spin alignment is largely dependent on halo formation time: if haloes form prior to entering filaments, they are less susceptible to spin swings thus their spin remains parallel aligned, whereas haloes that accrete significant mass upon entering filaments become orthogonally aligned. The above constitutes good evidence that spin alignment is an effective tracer of halo/LSS evolution, if distinct between cosmologies, could serve as a useful probe.      

Simulation-based investigations beyond spin-alignment have also proven useful in drawing distinctions between \lcdm{} and alternative models: \citet{Carlesi_14a} finds, using hydrodynamical simulations, that a self-interacting quintessence model provides a higher concentration of haloes within the LSS as compared to their fiducial \lcdm{} and other cosmologies compared. \citet{Carlesi_14b} find a weak coupling between the spin, triaxiality and virialisation and the cosmology dark sector types. Although \citet{Elahi_15} shows little to no difference between \lcdm{} and the coupled cosmologies judging by the lack of systematic differences between spin parameter and satellite alignment distributions across cosmologies. \citet{Smith_11} find that WDM model suppresses the halo mass function by 50$\%$ for masses 100 times the free-streaming mass scale. Recently it was found there are higher cluster abundances and lower void abundances within the \qcdm{} with respect to \lcdm{} \citep{Watts_17}. Moreover, differences were found between the densities of voids for each cosmology, being emptiest in \qcdm{} and densest within \lcdm{} \citep{Adermann_18}. They also suggest there is a potentially observable difference between the volume distribution of voids between \lcdm{} and CDE at low redshift. The differences in void/cluster statistics suggests distinct evolutionary paths are taken for non-standard cosmologies which may have a knock-on effect on spin-alignment evolution.

We aim to probe non-standard cosmologies using halo spin-LSS alignments and other halo statistics in order to reveal any cosmological signatures which may manifest as a result of the differing dark sector physics. This paper is structured as follows: Section \ref{nstdsec} will detail the non-standard cosmologies considered within this investigation. Section \ref{methsec} explains the methodology of LSS classification, the production of halo catalogues and alignment statistics. Then follows Section \ref{ressec} which elucidates the main findings of the investigation before a discussion of the results in Section \ref{discsec} and conclusion in Section \ref{concsec}. 

\begin{figure*}
\centering
\includegraphics[width=0.8\textwidth]{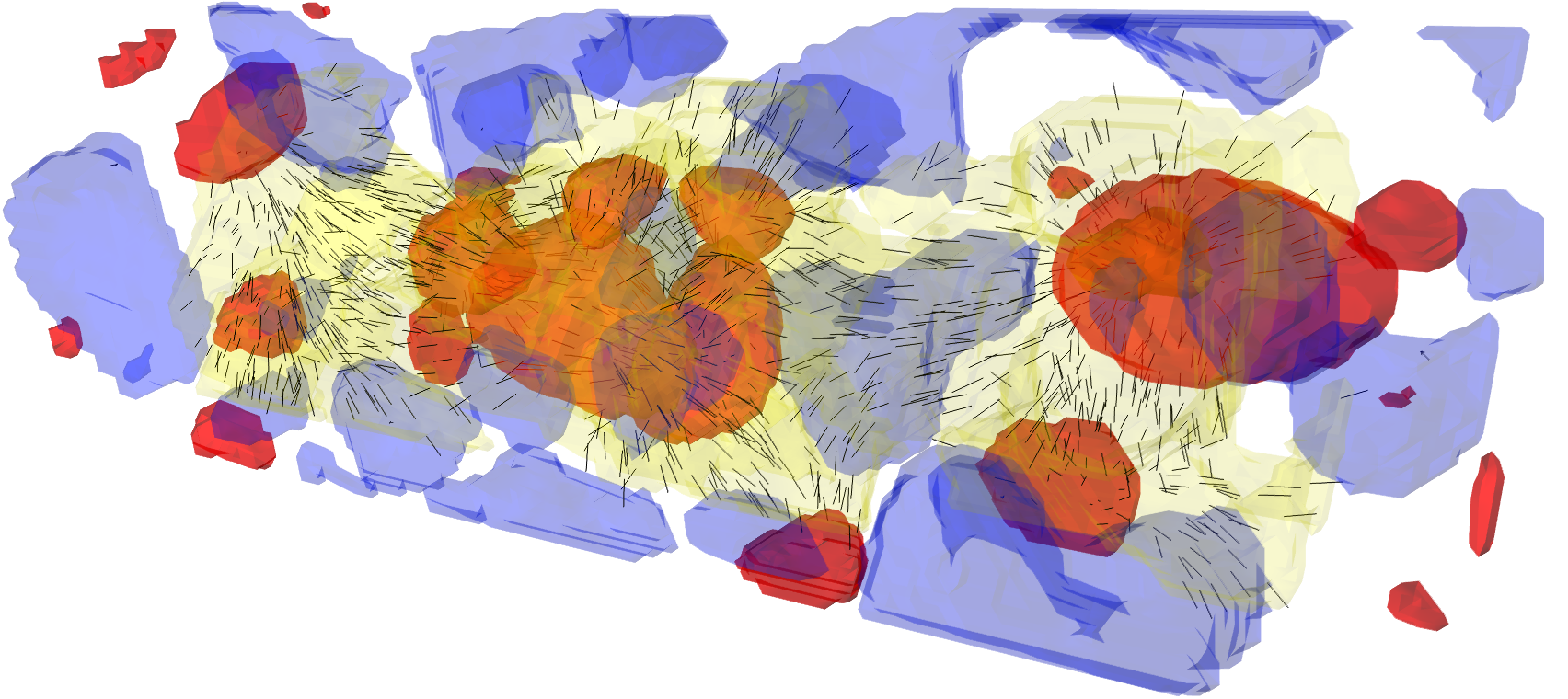}
\caption{Cylindrical cut-out of the \lcdmdm{} simulation where classified LSS are color-coded: red regions indicate clusters, yellow indicates filaments, and blue indicates voids. Sheets have been omitted for clarity, but occupy the white space in between the other LSS. Grey lines represent filament axes $\textbf{e}_{3}$ (tidal field direction). The Hessian Method allows the demarcation of the four main LSS types for the entire simulation volume. Also, it accurately traces the filament direction between clusters.}\label{figmayavi}
\end{figure*}

\section{Non-Standard Cosmological Models}\label{nstdsec}
We investigate three dark matter only simulations (labelled with subscript DM) including an uncoupled Quintessence model \qcdmdm{} and a Warm Dark Matter \lwdmdm{} model with fiducial model \lcdmdm{}. We also analyze a hydrodynamical simulation suite (with subscript DM+Gas) which includes three cosmologies: an uncoupled Quintessence model \qcdmg{} and two coupled models \cdeodmg{} and \cdeoodmg{} with fiducial \lcdmg{}.

\subsection{Uncoupled Quintessence (\qcdm{})}\label{nstdqcdm}
The general form of the Lagrangian which describes the scalar field of \qcdm{} model is represented by $\mathcal{L}$: 

\begin{ceqn}
\begin{equation}
\mathcal{L}=\int d^{4}x\sqrt[]{-g}\big(-\frac{1}{2}\partial_{\mu}\partial^{\mu}\phi+V(\phi)+m(\phi)\psi_{m}\bar{\psi}_{m}\big)\label{eqlagr},
\end{equation}
\end{ceqn}
including a kinetic term, potential term $V(\phi)$ and an interaction term $(\psi_{m})$ with dark matter.

The \qcdmdm{}/\qcdmg{} model includes no direct interaction between dark matter and the quintessence field, thus $m(\phi)=m_{0}$. But $\Lambda$ is replaced with a time-dependent, evolving scalar field $\phi$ (in units of Planck mass) whereby regions of the Universe have the opportunity to expand independently, thus alleviating the cosmological constant problem \citep{Joyce_15}. As for the potential term $V(\phi)$, this cosmology uses the Ratra-Peebles Potential  \citep{Ratra_88},

\begin{ceqn}
\begin{equation}
V(\phi)=V_{0}\phi^{-\alpha},\label{eqpot}
\end{equation}
\end{ceqn}
where $V_{0}$ and $\alpha$ are observationally fitted constants. This potential term is a contrived field potential, thus at late times the quintessence field dominates the energy budget of the universe, at early times it "tracks" the energy density of matter and radiation akin to observations \citep{Joyce_15}.

\subsection{Coupled Dark Energy (CDE)}\label{nstdcde}
We also investigate two coupled dark sector cosmologies, whereby the DM decay resolves the coincidence problem of \lcdm{}. The interaction is executed via a non-zero interaction term $m(\phi)=m_{0}e^{-\beta(\phi)\phi}$. We consider two coupled cosmologies which can be distinguished by their coupling terms of $\beta(\phi)=\beta_{0}=0.05$ (\cdeodmg{}) and $\beta(\phi)=\beta_{0}=0.099$ (\cdeoodmg{}). $\beta(\phi)$ are chosen to test the boundaries of allowed coupling with an effort to maximize any observational differences between the fiducial \citep{Pettorino_12}. Coupling allows for DM particles to decay into the DE scalar field $\phi$ resulting in an additional frictional force felt by the DM particles. This extra force ultimately effects the evolution of the density perturbation amplitude for DM, which in turn alters the baryon fraction within cluster-sized haloes \citep{Baldi_10}.

\subsection{Warm Dark Matter  ($\Lambda$WDM)}\label{nstdwdm}
\lcdm{} is extremely successful on large scales, thus WDM aims to merely modify the \lcdm{} model in a effort to dampen the small-scale structure of the Universe. \lwdmdm{} features dark matter particles which move at relativistic velocities, increasing the length of the free-streaming of particles, which smooths out over-densities and suppresses structure formation at scales smaller than the co-moving scale \citep{Bode_01}.
The free-streaming limit of the WDM particle has been confined via observations of the Lyman-alpha forest, to a lower limit mass $m_{WDM}\gtrapprox$ 3.3keV \citep{Viel_13}. Despite this lower limit, the mass we assign to our WDM model has particle energy $m_{WDM}$= 2keV, to exaggerate the cosmological effects. By multiplying the initial power spectrum by a transfer function, we truncate structure formation at the scale of 0.15 $h^{-1}$ Mpc \citep{Bode_01}.

\section{Methodology}\label{methsec} \FloatBarrier

\subsection{Cosmological Simulations}\label{methsims}
Before presenting our simulations, it is important to highlight the flexibility one has in the initial conditions when modelling non-standard cosmological models. Different cosmological observations constrain different parameters, for example the CMB constrains the amplitude of the matter power spectrum, and supernovae measure the expansion rate over cosmic time at moderately low redshift. In addition, observational data must be interpreted within the context of a model, which may be poorly constrained if only one dataset is used. Given this, cosmological parameters shared between models, such as the matter power spectrum normalization parameter $\sigma_8$ at z=0, may be well constrained within the context of \lcdm{} but not in an alternative model. This allows some flexibility in setting cosmological parameters for the non-standard cosmologies outlined in Section \ref{nstdsec}. 

For example, \citet{Baldi_12} studied some coupled models and allocated shared cosmological parameters based on estimates derived from CMB observations ($z\approx1100$) through a \lcdm{} lens. Another option, which we have opted for and was used by \citet{Carlesi_14a}, is one in which we use matter density parameters based on Planck CMB measurements constrained at $z=0$ in a \lcdm{} model and normalize each cosmology such that the amplitude $\sigma_8(z=0)$ is the same. Both approaches are valid. The flexibility means that differences seen in comparing different cosmologies may arise from different cosmological parameters rather than those arising from differences in  physics. 

Following the methodology of \citet{Elahi_15}, we use the \textsc{dark-gadget} N-body code to generate our cosmological simulations. \textsc{dark-gadget} is a modified version of \textsc{p-gadget}-2 (which is also a modified version of \textsc{gadget-2} see \citet{Springel_05,Carlesi_14a}), the key modification being the inclusion of a separate gravity tree to account for additional long-range forces and an evolving DM-particle mass to allow for decay. All simulations are initiated from $z$=100 with the same phases in their density perturbations, resulting in under-dense and over-dense regions in similar locations. This was done to eliminate random discrepancies when comparisons are made between cosmologies, thus any differences found between cosmologies are attributed to the differing amplitudes and dark sector physics. 

Initial conditions are produced using a uniform Cartesian grid along with the first-order Zel'dovich approximation using an altered version of the publicly available \textsc{n-genic} code. For non-standard cosmologies, the code requires full evolution of the scalar field $\phi$, the mass of DM-particles and the expansion history. In order to calculate the evolving linear power spectrum of the non-standard cosmologies along with the growth rate $\textit{f}\equiv d\ln D(a)/d\ln\textit{a}$, the publicly available \textsc{cmbeasy} \citep{Doran_05} is used, a Boltzmann code used to solve first-order Newtonian perturbation equations. The growth factors and expansion history calculated by \textsc{cmbeasy} are used to solve the particle displacements in the non-standard cosmologies. 

All cosmologies have $z=0$ parameters values of  (h,$\Omega_{m}$,$\Omega_{b}$,$\sigma_{8}$)=(0.67, 0.3175, 0.049, 0.83), consistent with \lcdm{} Planck data \citep{Plank_14b,Plank_16}. All cosmologies feature a 500$h^{-1}$Mpc sized box with $512^{3}$ particles ($2\times 512^{3}$ for DM+Gas simulations), where the particles (at z=0 for all cosmologies) have masses $m_{DM}=8.2\times 10^{10}h^{-1}M_{\odot}$ and $m_{DM}(m_{gas})=6.9(1.3)\times 10^{10}h^{-1}M_{\odot}$ for DM and DM+Gas suites, respectively. 
We ran simulations from $z$=100, producing 10 snapshots from 0$\leq$$z$$\leq$10. We look at 7 snapshots, beginning at $z$=2.98. We have limited our investigation to 0$\leq$$z$$\leq$2.98 as before this redshift there are not enough haloes to give robust statistics.

\subsection{Large-Scale Structure Classification}\label{methlss}

We classify the LSS for each cosmological simulation, at each snapshot taken from 0$\leq$$z$$\leq$2.98. We use the particle positions and masses acquired from snapshots to generate a density field. We utilize the Delaunay Tessellation Field Estimator (DTFE) in order to generate a fine-tuned density field. DTFE is an open source, C++ code \citep{Shaap_00,Weygaert_09,Cautun_11}. It is an adaptive method of density interpolation, in that it seeks out over-densities at the maximum possible resolution. It functions as follows:

1. DTFE creates Delaunay tetrahedra using the particle distribution.

2. Then, Voronoi cells are created from the tetrahedra, in which the density is then interpolated as a continuous field by using the volume of the cell along with the mass of each particle at its vertices.

Although the simplest method of generating a density field is by binning the positions of particles within a three-dimensional grid, this method leads to unphysical discontinuities and shot noise at high resolution, which DTFE alleviates. We use the 3-dimensional density field $\rho(\textbf{x})$ produced from DTFE to form the Hessian matrix,
\begin{ceqn}
\begin{equation}
\textbf{\textit{H}}_{\alpha \beta}=
\begin{bmatrix}
    \frac{\partial^{2}\rho(\textbf{\textit{x}})}{\partial \textit{x}_{x}^{2}} & \frac{\partial^{2}\rho(\textbf{\textit{x}})}{\partial \textit{x}_{y}\partial \textit{x}_{x}} & \frac{\partial^{2}\rho(\textbf{\textit{x}})}{\partial \textit{x}_{z}\partial \textit{x}_{x}}\\
    \frac{\partial^{2}\rho(\textbf{\textit{x}})}{\partial \textit{x}_{x}\partial \textit{x}_{y}} & \frac{\partial^{2}\rho(\textbf{\textit{x}})}{\partial \textit{x}_{y}^{2}} & \frac{\partial^{2}\rho(\textbf{\textit{x}})}{\partial \textit{x}_{z}\partial \textit{x}_{y}}\\
    \frac{\partial^{2}\rho(\textbf{\textit{x}})}{\partial \textit{x}_{x}\partial \textit{x}_{z}} & \frac{\partial^{2}\rho(\textbf{\textit{x}})}{\partial \textit{x}_{y}\partial \textit{x}_{z}} & \frac{\partial^{2}\rho(\textbf{\textit{x}})}{\partial \textit{x}_{z}^{2}}\label{eqhessmat}
\end{bmatrix}
\end{equation}
\end{ceqn}
whereby $\alpha, \beta=x,y,z$. It is a second-order partial derivative of $\rho(\textbf{x})$ for each of the nine unique directions (although only six elements are unique due to directional symmetries). This matrix encapsulates the local curvature for each voxel in $\rho(\textbf{x})$ and allows us to characterise the LSS by its eigen-pairs (as demonstrated in \citet{Calvo_07,Hahn_07}). Prior to forming the Hessian matrix, smoothing the density field is paramount to produce a field that probes density gradients on appropriate scales. We convolve the raw density field $\rho(\textbf{x})$ with a 3-dimensional Gaussian distribution kernel $G_{s}$,

\begin{ceqn}
\begin{equation}
G_{s}=\frac{1}{(2\pi\sigma^{2}_s)^{3/2}}e^{\Big(-\frac{(x^{2}+y^{2}+z^{2})}{2\sigma^{2}_{s}}\Big)}
\label{eqgaus}
\end{equation}
\end{ceqn}
for multiple smoothing scale widths ($\sigma_s$=2,3.5 and 5 Mpc/$h$) as increasing the smoothing scale allows us to investigate the alignment of halo spin with structure at different scales. \citet{Trowland_13} demonstrated the utility of multiple smoothing scales by tracking the evolution of alignments across multiple smoothing scales and discovered evidence of filament thickening over time. We calculate the Hessian matrix in k-space and Fourier transform to determine $\textbf{\textit{H}}_{\alpha \beta}$ for each voxel in the simulation volume.

\begin{table}
\centering
\begin{tabular}{|c|c|c|c|}
	\hline
	LSS Type & $\lambda_{1}$ & $\lambda_{2}$ & $\lambda_{3}$ \\
	\hline
    Cluster & $<0$ & $<0$ & $<0$ \\
	\hline
    Filament & $<0$ & $<0$ & $>0$ \\
	\hline
    Sheet & $<0$ & $>0$ & $>0$  \\
	\hline
    Void & $>0$ & $>0$ & $>0$  \\
	\hline   
    
\end{tabular}
\caption{\label{tabeigpairs}The Hessian Method is used to generate eigenpairs which comprise this classification scheme. The LSS are classified depending on the combination of eigenvalues ($\lambda_{1}$, $\lambda_{2}$ and $\lambda_{3}$). The LSS axes are represented by the last collapse direction (the eigenvector $\textbf{e}_{3}$) regardless of the LSS type.} 
\end{table}

Each voxel is characterised as a cluster, filament, wall or void based on the ordered eigenvalue as per table 1, with direction of last collapse defined by $\mathbf{e}_3$, which will be used to represent the axes of all LSS. Many works have studied the alignment of halo spin with LSS, finding it is not exclusive to filamentary structure, but is also seen within sheets and clusters, where there is a general alignment between halo spin and the $\textbf{e}_{3}$ (initial intermediate tidal tensor) axes, also known as the axes of slowest collapse \cite[e.g.][]{Libeskind_12,Dubois_14,Calvo_14,Kang_15,Wang_17,Veena_18}. Furthermore, \citet{Bond_96,Codis_12,Pichon_16} state that the spin evolution of haloes are part in parcel of the pancaking effect whereby the fastest collapsing axis $\textbf{e}_{1}$ causes mergers and accretion along this axis and explains the alignments found. 

With knowledge of the LSS type and direction of each voxel within our simulations, we are able to classify each structure, as seen in Figure \ref{figmayavi}. This figure shows a cylindrical sample cut-out of an arbitrary snapshot. Clusters are represented as red blobs, filaments by the yellow iso-surface and voids as blue blobs. Sheets have been omitted from the visualization for clarity, but would occupy the volumes between the mentioned LSS. This cut-out centres around a filament which is strung by a few clusters and surrounded by sheets and voids. The small grey lines delineate the filament axes ($\textit{e}_{3}$). We can see they align with the length of the yellow filament, pointing towards the clusters where the bulk flow of matter is directed.  

\subsection{Halo catalogues}\label{methhalos}
The very same particles which formed the density field from each snapshot are used for VELOCI\textsc{raptor} (A.K.A STructure Finder), a dark matter halo classifier \citep{Elahi_11,Elahi_19a}. VELOCI\textsc{raptor} is a sub(halo) finder which works in a two-step process:

1. Haloes are first identified using a three-dimensional Friends-of-Friends (3DFoF) algorithm pruned for any artificial particle bridges using a 6DFoF and the velocity dispersion of the FoF group.

2. Then substructure is found by inspecting the dynamics, such as velocity distribution of particles and their distinctness with the halo environment. The substructure is linked via a phase-space FoF algorithm.
\begin{figure}
\centering
\includegraphics[width=0.45\textwidth]{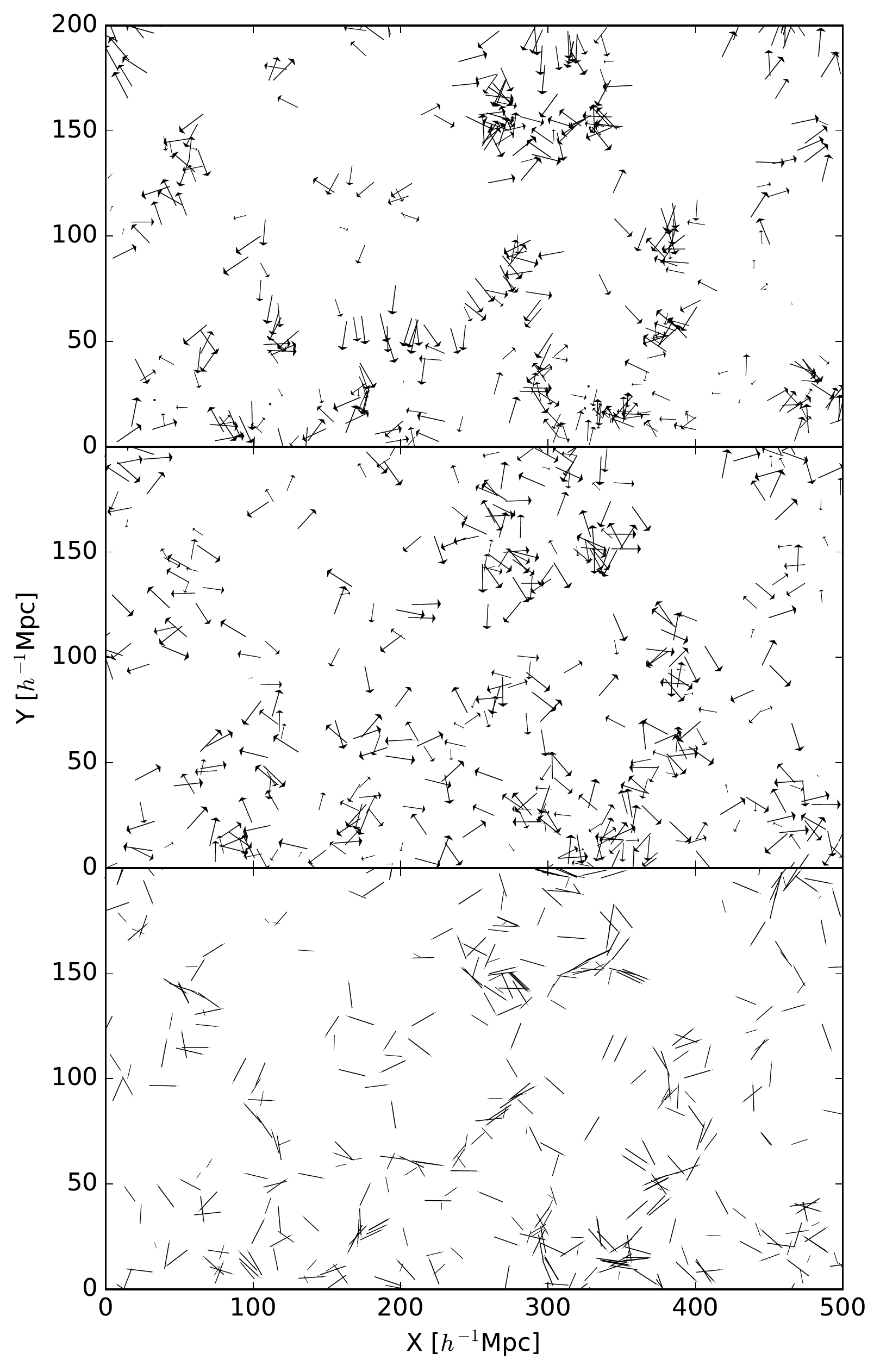}
\caption{Three arbitrary 500 $\times$ 200 $\times$ 40 $\textit{h}^{-1}$Mpc slices of haloes and coinciding filament axes. Top panel: halo velocity vectors at halo positions. Middle panel: halo spin vectors. Bottom panel: filament axes at halo positions. We measure alignments by taking the dot product between halo vectors (spin and velocity) and the filament axes which coincide with halo positions.}\label{figvecslc}
\end{figure}
VELOCI\textsc{raptor} also generates the spins of the catalog haloes. The spin of each halo is calculated by including all the associated particles ($N$) for each halo,
\begin{ceqn}
\begin{equation}
\textbf{J}=\sum_{i=0}^{N}\textbf{r}_i\times( m_i\textbf{v}_i )\label{eqspin}
\end{equation}
\end{ceqn}
where $\textbf{r}_i,m_i$ and $\textbf{v}_i$ are the particle radius from halo centre, mass and velocity, respectively. We filter out haloes with $\leq$100 particles as the spin measurements are unreliable below this particle quantity threshold. This limits our low mass haloes to beyond the parallel alignment mass range (typically of the order $10^{12} \Msun{}$), thus we are unable to detect parallel alignment.

We construct a halo merger-tree to track haloes and identify their progenitors through simulation snapshots. We implement \textsc{treefrog}, a code which is part of the VELOCI\textsc{raptor} package as per \citet{Elahi_19b}. \textsc{treefrog} works by tracking particles through the simulation snapshots and identifies optimal halo links via the merit function,  

\begin{ceqn}
\begin{equation}
\mathcal{N}_{A_{i}B_{j}}=\frac{N_{{A_{i}\cap B_{j}}}^{2}}{N_{A_{i}}N_{B_{j}}}    \label{eqmerit}
\end{equation}
\end{ceqn}
where $N_{{A_{i}\cap B_{j}}}$ is the number of shared particles between catalog objects and $N_{A_{i}}$ and $N_{B_{j}}$ are the total number of particles for each corresponding object within the catalogues. Catalogues $A$ and $B$ are cross-matched by finding each object within catalog $A$ and each object within catalog $B$ which maximizes the merit $\mathcal{N}_{A_{i}B_{j}}$. We link our haloes across catalogues by choosing the candidate haloes which present the largest merit (following the main branch of the merger tree) value across every successive catalog between 0$\leq$$z$$\leq$2.98 as this is generally a robust method for identifying matches. There are other methods, see \citet{Li_08} for examples.

\begin{figure}
\centering
\includegraphics[width=0.42\textwidth]{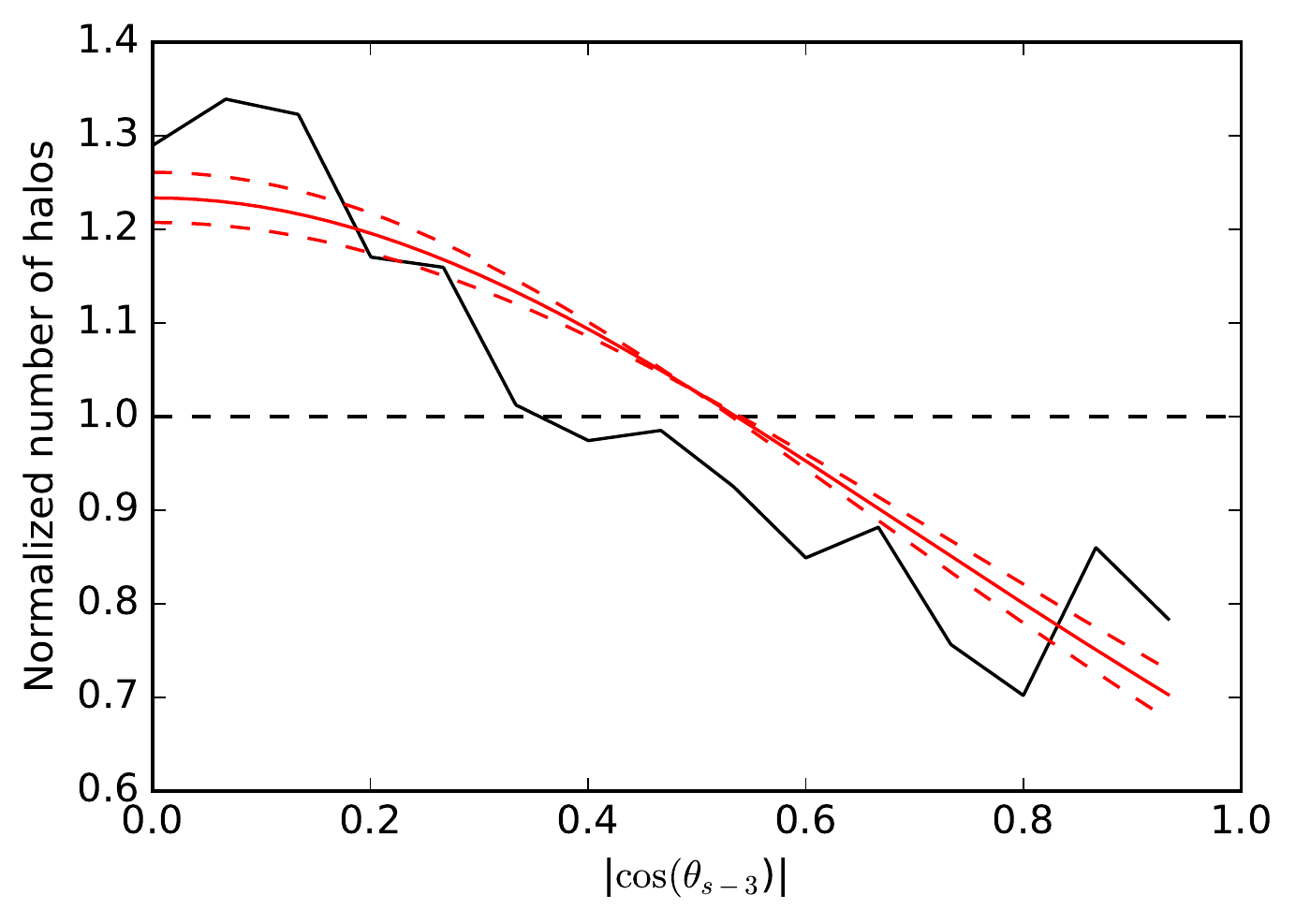}
\caption{An example of a normalized dot product distribution wrt to filaments (black solid line) fitted with the PDF model (Equation \ref{eqmod}) resulting in a $c=0.23\pm0.02$ which is represented by the red solid line. The red dotted lines show the 1$\sigma$ models. We find a skew towards lower values of |$\cos{\theta_{s - 3}}$| indicating an orthogonal alignment, where a random distribution of alignments is represented by the flat dotted line.}\label{figmodfit}
\end{figure}

\subsection{Alignment Statistics}\label{methalgnstat}
Figure \ref{figvecslc} displays a 40$\textit{h}^{-1}$Mpc slice of a simulation sample where we have filtered out haloes which reside within filaments, using the eigenvalue criteria for a filament within Table \ref{tabeigpairs}. The velocities of haloes (top panel), spins of the same haloes (middle panel) with coinciding filament axes (bottom panel) can be aligned. In order to calculate alignment between halo spin and filament axes we take the dot product of the unit spin vector \textbf{J} with the filament axes unit vector $\textbf{e}_{3}$:

\begin{ceqn}
\begin{equation}
\cos{\theta_{s - 3}}=\bigg|\frac{\textbf{J}\cdot \textbf{e}_{3}}{|\textbf{J}|}\bigg|.\label{eqdp}
\end{equation}
\end{ceqn}
Since all we require is the alignment, the absolute value of the dot product is taken which leaves us with the alignment $\theta$:[0\textdegree, 90\textdegree], although alignment is typically represented with $\cos{\theta}:[0,1]$. 

There are two methods by which alignment statistics are represented, the first and most common is by simply taking the mean alignment $\langle \cos{\theta}\rangle$ of halos binned by mass. Here we calculate the distribution of the mean using bootstrap re-sampling to determine the most likely value along with the 1$\sigma$ error. We also use the raw dot-product distribution with a Probability Density Function (PDF) (model) derived by \citet{Lee_11} from TTT,

\begin{ceqn}
\begin{equation}
\textit{P}(\cos{\theta})=(1-c)\sqrt{1+\frac{c}{2}}\Big[1-c\Big(1-\frac{3}{2}\cos^{2}{\theta}\Big)\Big]^{-3/2}. \label{eqmod}
\end{equation}
\end{ceqn}
The model takes into account the unit spin vector of structures and the tidal field in order to quantify the degree of alignment. It is fitted to the distribution of the alignment between halo spin and filament axes. The value of the correlation parameter $c$ represents the strength of alignment. A positive $c$ represents an orthogonal alignment and would agree with TTT predictions, but a negative $c$ which indicated a parallel alignment to the tidal field is not predicted by TTT. The spin may also flip from being parallel to orthogonal to the tidal field, especially for high mass haloes, due to mergers, accretion and formation time/entering time of haloes \citep{Wang_17}. Figure \ref{figmodfit} shows an example of the alignment to filaments. The PDF model was fitted to the distribution by constructing a grid in the range $c$:[-1,1] with a $10^{-4}$ interval to form the likelihood function. The uncertainty of $c$ represents the values at $\sim\pm34\%$ of the area under the likelihood function (i.e 1$\sigma$ for a Gaussian distribution) from the mean value.

\section{Results}\label{ressec} 

Upon basic inspection of halo quantities for each cosmology, we find on average 10$\%$ more haloes to be within \qcdmdm{} at low redshift, compared with \lcdmdm{}. For the DM+Gas simulations we find that coupled cosmologies (\cdeodmg{} and \cdeoodmg{}) have the largest halo quantity discrepancy with \lcdmg{} of an average 10$\%$ at low redshift. Generally, halo quantities of masses $\leq 10^{13.5}\Msun$ are discrepant across cosmologies and \lcdm{} consistently features the lowest quantity of haloes at all redshifts, for both DM and DM+Gas simulation suites. Halo quantity discrepancies across cosmologies may influence other statistical comparisons made throughout, although this should be encapsulated in signal errors.


\begin{figure*}
\centering
\includegraphics[width=0.9\textwidth]{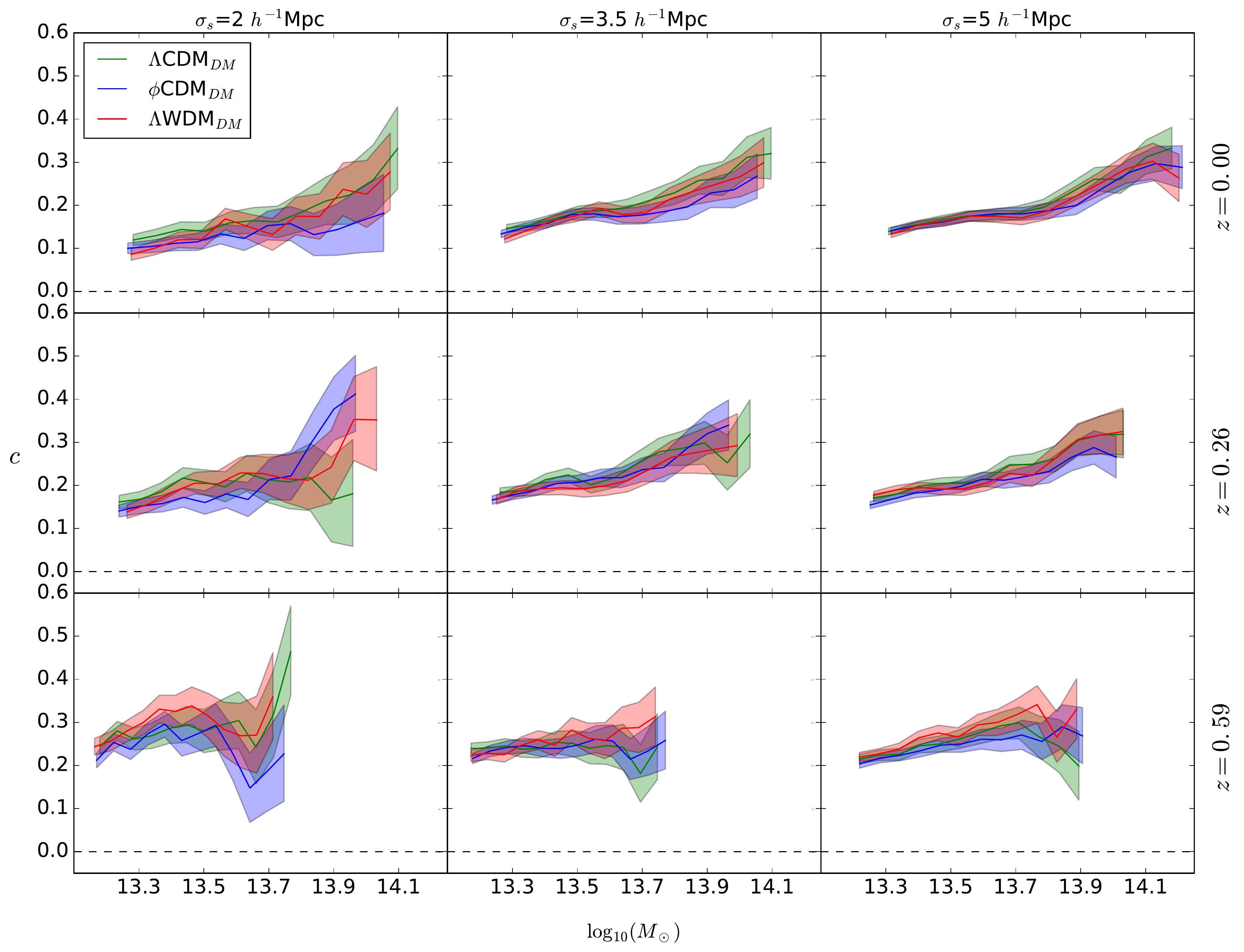}
\caption{Model-fitted halo spin-filament alignments organized for Rows: top to bottom as $z$=0, 0.26 and 0.59. Columns: left to right shows 2, 3.5 and 5$\textit{h}^{-1}$Mpc smoothing scales. Colour coded for each DM cosmology. All DM cosmologies show a positive correlation between $c$ and halo mass, regardless of redshift or smoothing scale.}\label{figspinfildm}
\end{figure*}

\begin{figure*}
\centering
\includegraphics[width=0.93\textwidth]{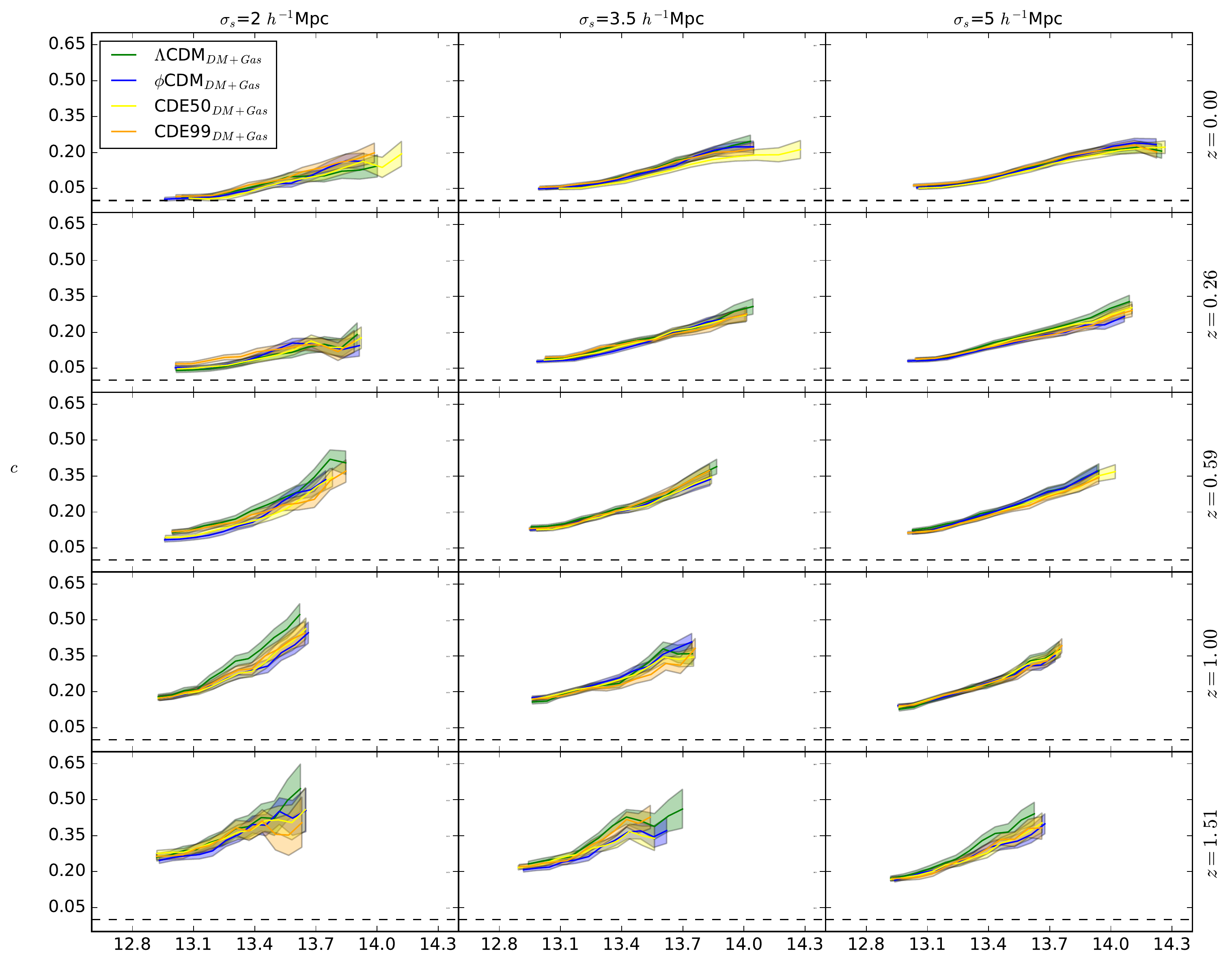}
\caption{Similarly defined as Figure \ref{figspinfildm}. DM+Gas simulations, colour coded by Cosmology for all smoothing scales and redshifts. The same mass-scale dependency is found for virtually all smoothing scales and at all $z$, as per the DM alignments. In addition, all alignments are consistent across cosmologies.}\label{figspinfildmg}
\end{figure*}

\begin{figure*}
\centering
\includegraphics[width=0.9\textwidth]{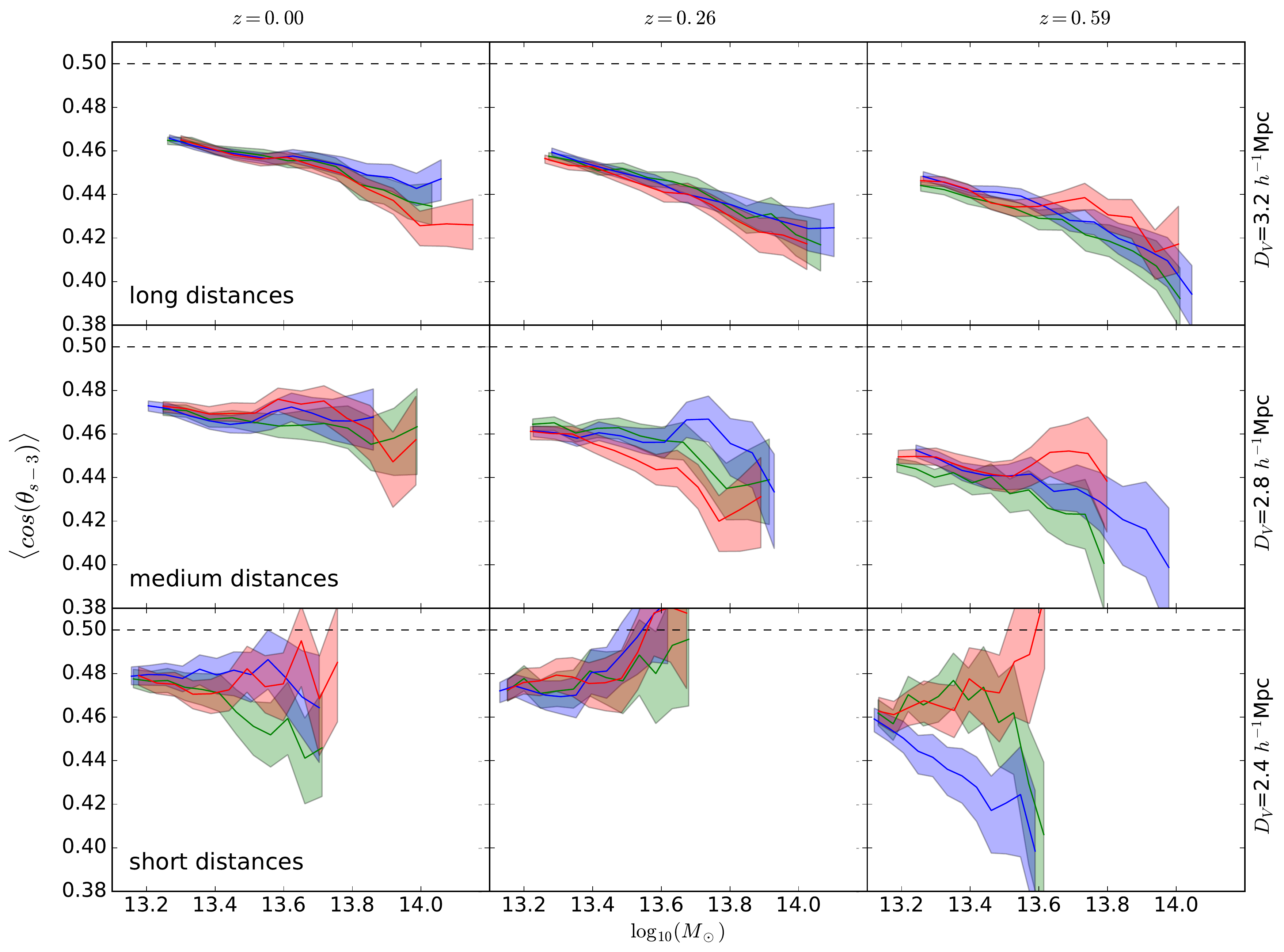}
\caption{DM simulations model-fitted haloes spin-LSS alignment within void regions organized for Rows: top to bottom as $D_{V}$=3.2,2.8 and 2.4 $\textit{h}^{-1}$Mpc. Columns: left to right shows $z$=0, 0.26 and 0.59. Color coded as per Figure \ref{figspinfildm}. Haloes on the outskirts of voids show no signs of cosmological signatures.}\label{figvsspinlssdm}
\end{figure*}

\begin{figure*}
\centering 
\includegraphics[width=0.9\textwidth]{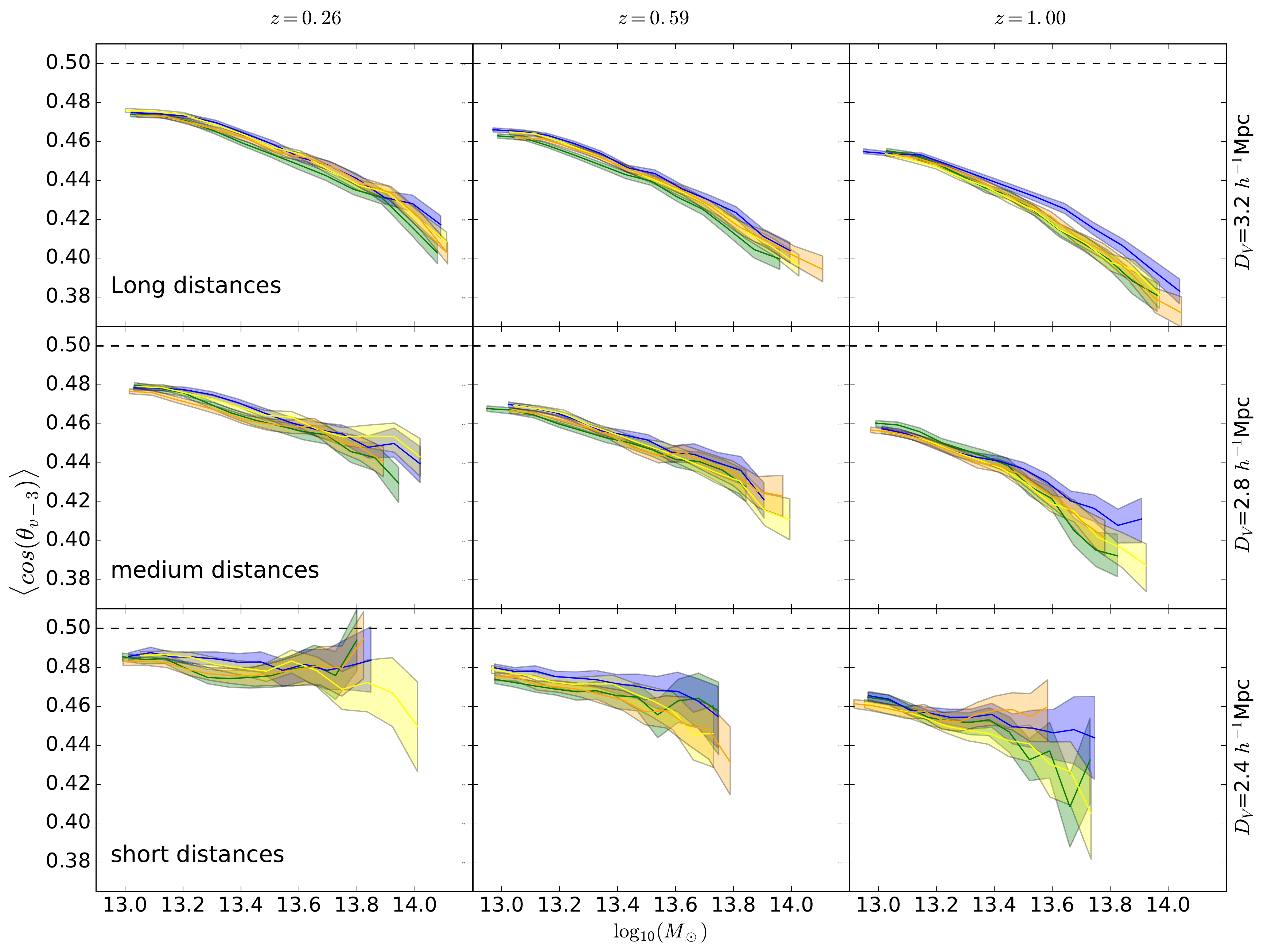}
\caption{DM+Gas simulations model-fitted haloes spin-LSS alignment within void regions organized for Rows: top to bottom as $D_{V}$=3.2,2.8 and 2.4 $\textit{h}^{-1}$Mpc. Columns: left to right shows $z$=0, 0.26 and 0.59. Color coded as per Figure \ref{figspinfildmg}. \qcdmg{} features a consistently weaker orthogonal alignment, suggesting perhaps an effect of matter evacuating out of voids in distinction with other cosmologies.}\label{figvsspinlssdmg}
\end{figure*}

\begin{figure}
\centering
\includegraphics[width=0.47\textwidth]{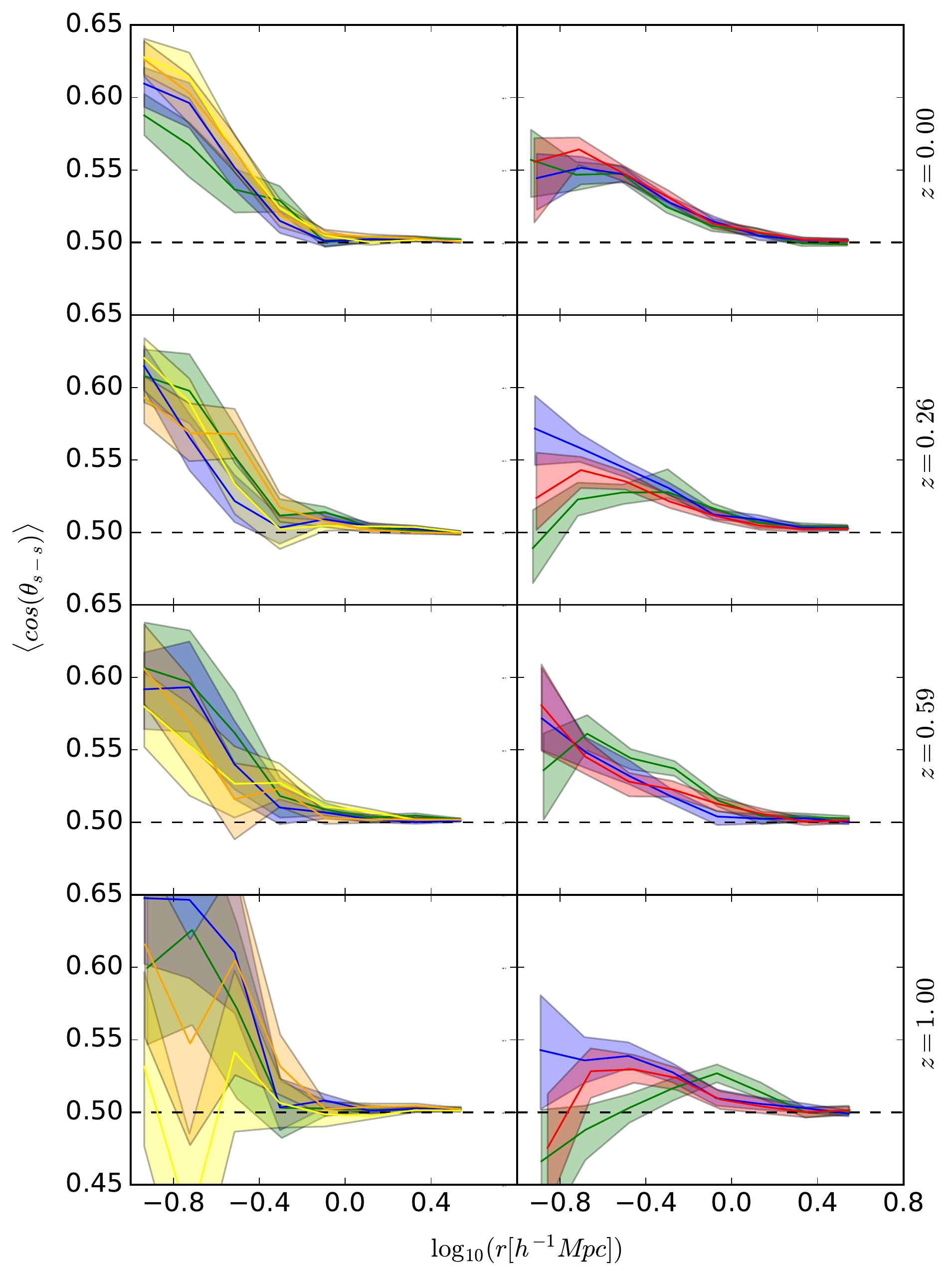}
\caption{Haloes' average spin-spin alignments as a function of separation distance. Values above the horizontal dotted line represent a parallel alignment, and orthogonal alignment for below 0.5. DM+Gas simulations (left column, color coded as per Figure \ref{figspinfildmg}) and DM simulations (right column, color coded as per Figure \ref{figspinfildm}) show strong parallel alignments at close distances and low redshift, but fade out at high redshift and large separation between haloes. Cosmological signatures are also missing at these small-scales. }\label{fighhspin}
\end{figure}


\subsection{Spin Alignment}\label{reslsa}
TTT predicts that proto-haloes are most susceptible to spin-up, and post-formation their spins are less likely to be disturbed. Given that each non-standard cosmology's tidal field may differ by a perceivable degree, spin-up of haloes may also differ in turn. This allows spin-filament alignments, as a function of halo mass for DM simulations, in Figure \ref{figspinfildm} and DM+Gas in Figure \ref{figspinfildmg} to detect such cosmological differences.

Both figures are generated in the same manner: As per Section \ref{methalgnstat}, we filter out all haloes (including subhaloes) within filament regions and take the dot product between their spin and coinciding filament axes. Then, instead of binning haloes as a function of mass in equal interval mass bins, we overlap bins (in what is known as a moving bin) so that each bin is 90$\%$ overlapped by its successive bin. The outcome is each bin is strengthened in quantity as haloes may appear within multiple bins, reducing the error of each data point. By using a moving bin the 1$\sigma$ error will be correlated for adjacent bins, but it does not negate any systematic differences between alignments that may arise as a result. For each mass bin of haloes we fit the model (Equation \ref{eqmod}) to the dot product distribution, as per example in Figure \ref{figmodfit}. We take a total of 12 mass bins per signal in Figures \ref{figspinfildm} and \ref{figspinfildmg} where $c$ is plotted at the centre of each bin interval. 

Figures \ref{figspinfildm} and \ref{figspinfildmg} display the alignments for all three smoothing scales, ordered as 2, 3.5 and 5$\textit{h}^{-1}$Mpc from left to right columns, at the given $z$ which proceeds down the rows as computed in the range $z$=0-2.98,  although figures display until $z$=0.59 and $z$=1.51 for DM and DM+Gas simulations, respectively. Given the higher the value of $c$, the more orthogonally aligned the haloes are with filament axes, we see there is a mass dependency whereby higher mass haloes have stronger orthogonal alignment, for all simulations. We also find there is a smoothing scale dependency for the DM+Gas simulations: in Figure \ref{figspinfildmg} from $z$=1.00-1.51 the $\sigma_{s}$=2$\textit{h}^{-1}$Mpc column shows the strongest orthogonal alignment out of the other smoothing scales, then a transition occurs at $z$=0.59 and continues until present day where the $\sigma_{s}$=5$\textit{h}^{-1}$Mpc shows the stronger alignments. 

We reiterate: only haloes with more than 100 particles are taken into account when calculating the alignments. This cut-off rules out low mass haloes, thus leaving only high mass haloes with a positive $c$ (not including eccentric dips below $c$=0 line which are attributed to scarcity of haloes within affected bins). DM + Gas simulations have twice the number of particles, leading to a significant increase in quantity of haloes at lower mass scales. Although parallel alignment (negative $c$ values) is not seen, the alignment uncertainties are significantly reduced across the redshift range. 

\begin{figure*}
\centering
\includegraphics[width=0.8\textwidth]{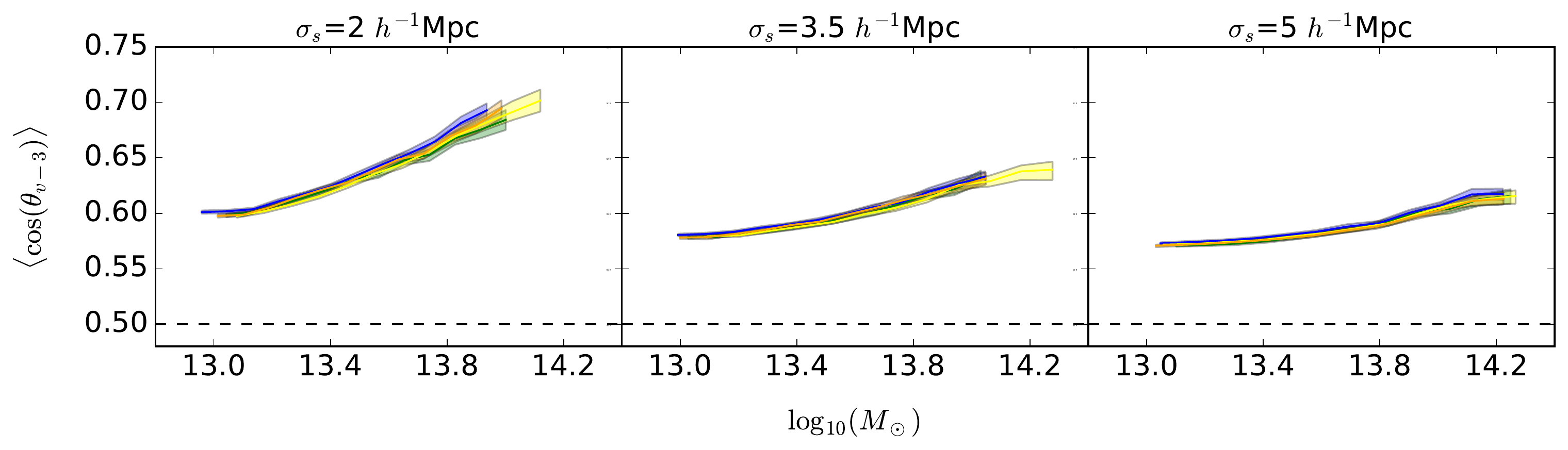}
\caption{Present day DM+Gas simulations, halo velocity-filament alignment as shown for $\sigma_{s}$=2, 3.5 and 5$\textit{h}^{-1}$Mpc smoothing, from left to right. Color coded as per Figure \ref{figspinfildmg}. Similar to Figure \ref{fighhspin} in that random alignment falls at 0.5, parallel and orthogonal alignment falls above and below, respectively. All signals show a strong parallel alignment peaking at the 2$\textit{h}^{-1}$Mpc smoothing scale, which suggests that haloes are streaming down filaments, with high mass haloes having the strongest alignment is also suggestive of mergers and accretion along the direction of flow.}\label{figvelfildmg}
\end{figure*}

Given halo spins are less likely to be disturbed in low density regions, we filter out haloes which reside on the outskirts of voids. We demarcate void regions as simply a distance threshold from the edges of voids to a particular distance beyond the void's edge. We have determined 3 relevant distance ranges: $D_{V}$=2.4,2.8 and 3.2 $\textit{h}^{-1}$Mpc from void's edge (where voids are classified within the $\sigma$=2$\textit{h}^{-1}$Mpc smoothing mask), nicknamed short, medium and long distances respectively shown in Figure \ref{figvsspinlssdm}. Shorter distances than 2.4$\textit{h}^{-1}$Mpc result in insufficient haloes for a statistical comparison between cosmologies, and beyond 2.8$\textit{h}^{-1}$Mpc result in signals akin to high density regions. Void region signals in Figures \ref{figvsspinlssdm} and \ref{figvsspinlssdmg} are organized by the three distances from voids for each row, as annotated. The signals are calculated via the first method stipulated in Section \ref{methalgnstat}, with 12 moving bins per signal. The DM+Gas simulations show consistent discrepancies between \qcdm{} and the fiducial where \lcdm{} haloes at all mass scales have a stronger alignment. This difference in alignment strength is at a maximum for the $z$=0.59 column, at mid-range masses. Within the DM simulations for void regions (Figure \ref{figvsspinlssdm}) the signals are more discrepant across cosmologies as compared with spin-filament alignment. This may be partially due to the reduced quantity of haloes within each mass bin, increasing the error but also the fluctuations between cosmologies.

Cosmological signatures are a second order effect, thus we investigate spin alignment on a smaller and more sensitive scale, between neighboring haloes, which may reveal unique signatures of non-standard cosmologies. This form of alignment has been found within simulations \citep{Bailin_05,Trowland_13} and observations \citep{Pen_00,Slosar_09} but is also predicted by TTT for haloes within proximity. In order to produce halo spin-spin alignment, haloes are paired as a function of distance $r$: we take the alignment of a halo at position $x$ in $\textbf{J}_{x}$ with its pair separated by $r$ in $\textbf{J}_{x+r}$ via the dot product method as per Section \ref{methalgnstat},

\begin{ceqn}
\begin{equation}
\cos{\theta_{s - s}}=\bigg|\frac{\textbf{J}_{x}\cdot \textbf{J}_{x+r}}{|\textbf{J}_{x}| |\textbf{J}_{x+r}|}\bigg|\label{eqdpss}
\end{equation}
\end{ceqn}
The process to obtain Figure \ref{fighhspin} is identical to Section \ref{methalgnstat} thereafter. There are 12 moving bins (with 90$\%$ overlap) for each signal for both DM (right column) and DM+Gas (left column) simulations. Importantly, we only take halo-halo and halo-subhalo pairs. Subhalo-subhalo alignment is complex as they have short time-scale interactions in a highly nonlinear environment with the tidal field of the halo. The redshift is organised by rows, ranging from $z$=0.00-1.00 from top to bottom row. In searching for halo pairs, we restricted the distance range $r$ between pairs to be $-0.8<\log_{10}{[r/(\textit{h}^{-1}Mpc)]}<0.8$, as from $\log_{10}{[r/(\textit{h}^{-1}Mpc)]}=0.5$ alignment are stable on 0.5, constituting a random alignment and below $\log_{10}{[r/(\textit{h}^{-1}Mpc)]}=-0.8$ alignments become too noisy due to scarcity of haloes. 

Values residing above 0.5 constitute a parallel alignment within Figure \ref{fighhspin}. DM simulations peak in their parallel alignment for haloes at distances of $\log_{10}{[r/(\textit{h}^{-1}Mpc)]}=-0.6$ with no significant differences across cosmologies. For DM+Gas simulations we find that results are noisier due to the reduction in the number of haloes within each bin but statistically significant parallel alignment is noticeable at low redshift. From low to high redshift it is found that parallel alignments flatten out progressively to random alignment. 

\begin{figure}
\centering
\includegraphics[width=0.42\textwidth]{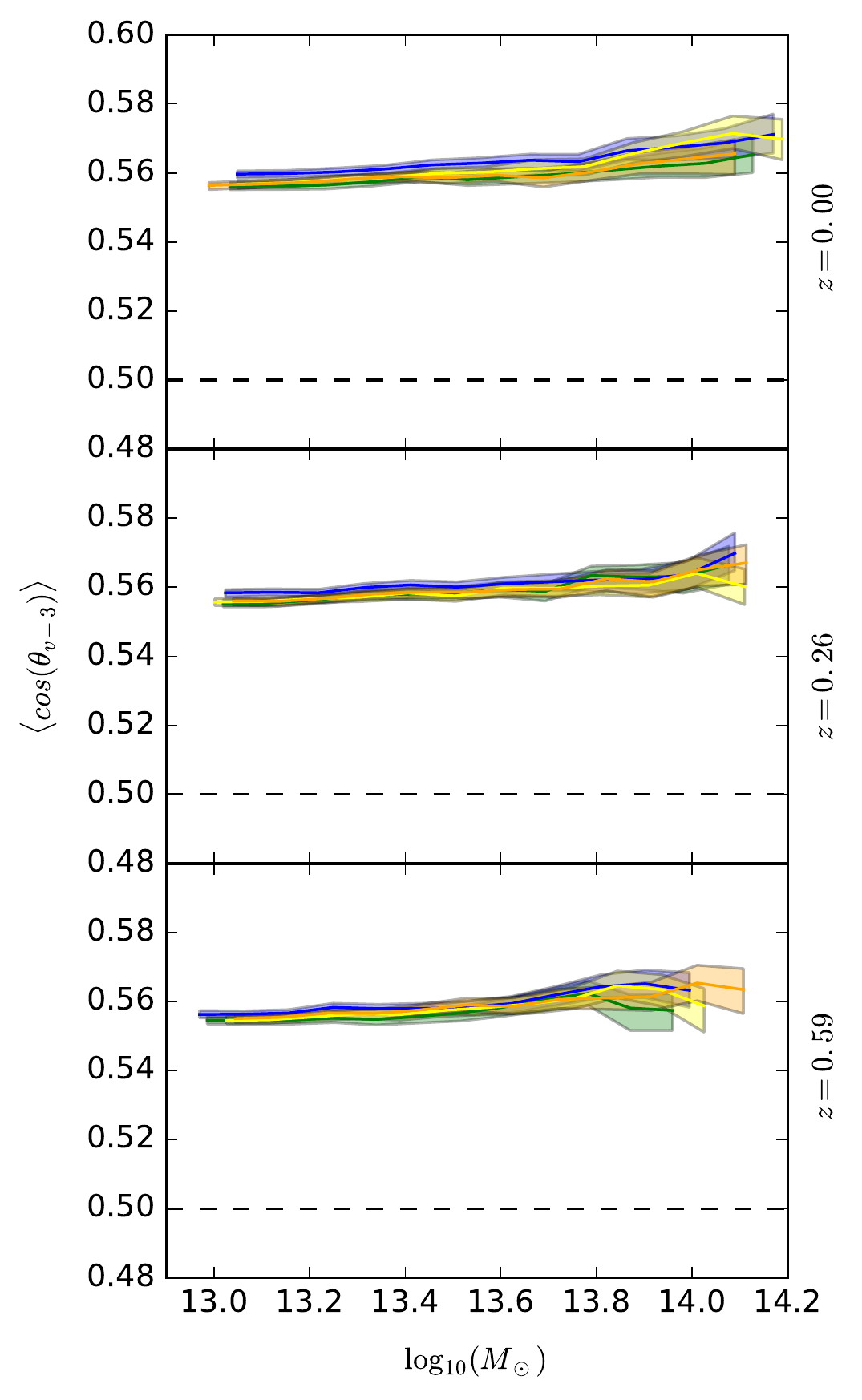}
\caption{Velocity-LSS alignment for haloes on the outskirts of voids at $D_{V}$=3.2$\textit{h}^{-1}$Mpc. Color coded as per Figure \ref{figspinfildmg}. The signals are not mass dependent although still maintain a parallel alignment across redshift.}\label{figvsvellssdmg}
\end{figure}

\subsection{Movement and Migration}\label{resmm}

From Section \ref{intro} we stated the differences found between cosmologies by \citet{Adermann_18} were that volumes of voids were found to be smaller yet emptier in the \qcdm{} cosmology. This research was conducted using the same DM+Gas simulation suite we analyze throughout this paper. Moreover, \citet{Watts_17} uses the same DM simulation suite and found that there are higher cluster abundances and lower void abundances within \qcdm{} also. Based upon these previous findings, we compare the velocities and migration patterns of haloes to see whether there is a mass flow discrepancy across cosmologies.

Figure \ref{figvelfildmg} displays the $z$=0 halo velocity-filament alignment $\langle\cos(\theta_{v -3})\rangle$, calculated by taking the mean of each mass bin as in Section \ref{methalgnstat}. We find there is a strong mass-dependence whereby higher mass haloes display a stronger parallel alignment. We also find the smoothing scale dependence of low smoothing scale resulting in slightly weaker parallel alignment that carry on from low to high redshift, although we omit signals from $z \geq$0.26 for brevity. We also looked at the velocity-LSS alignment (not included within this paper), but did not filter out haloes by filament regions as we did in Figure \ref{figvelfildmg}, rather we included all haloes. We found trends virtually mimicking Figure \ref{figvsvellssdmg} in that they are not mass dependant nor are they smoothing scale dependent. We omit such figures from our results as they resemble Figure \ref{figvsvellssdmg} which are in fact halo velocity-LSS alignment for haloes filtered for $D_{v}$=3.2$\textit{h}^{-1}$Mpc (long distance) void regions. Thus mass-dependence of velocity-LSS alignment is unique to filament regions only. Note however we do not fit the model (Equation \ref{eqmod}) to the alignments as it is TTT derived for filament axes only, but we take the mean and use bootstrap resampling for the error, although it bears no significance on comparisons made between model-fitted results as Figure \ref{figvsvellssdmg} signals would be mass-independent regardless.

\begin{figure}
\centering
\includegraphics[width=0.5\textwidth]{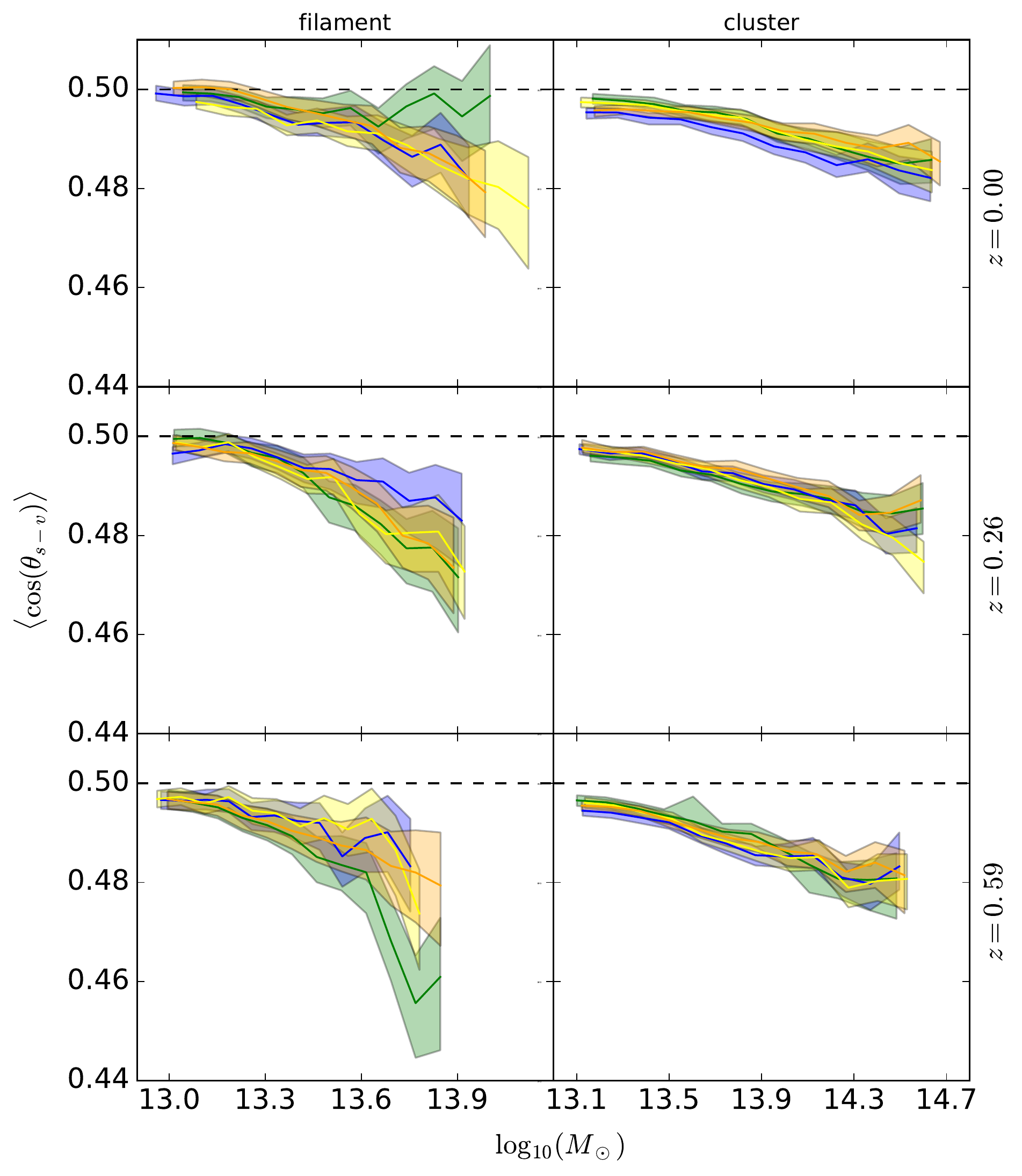}
\caption{Halo spin-velocity alignment signals for $z$=0, 0.26 and 0.59 from top to bottom. Color coded as per Figure \ref{figspinfildmg}. There is a weak orthogonal alignment between a halo's spin and velocity vectors for haloes within filaments (left column) and clusters (right column).}\label{figspinveldmg}
\end{figure}

We also take the dot product of the spin-velocity $\langle\cos(\theta_{s -v})\rangle$ of filament haloes (left column) and cluster haloes (right column) displayed in Figure \ref{figspinveldmg}. Since the alignment of halo spin is largely orthogonal(as per Figures \ref{figspinfildm} and \ref{figspinfildmg}), being more orthogonal for higher masses, and the velocity is increasingly parallel for higher mass, one would assume velocity-spin alignment would be strongly orthogonal. But we find in Figure \ref{figspinveldmg} there is only a weak orthogonal alignment which strengthens mildly at the increase of halo mass for filament-haloes and moreso for cluster-haloes. There is also a slight increase in the strength of orthogonal alignment as the redshift increases for both columns, which resembles the spin-filament alignment trend found within Figure \ref{figspinfildmg}.

\begin{figure}
\centering
\includegraphics[width=0.35\textwidth]{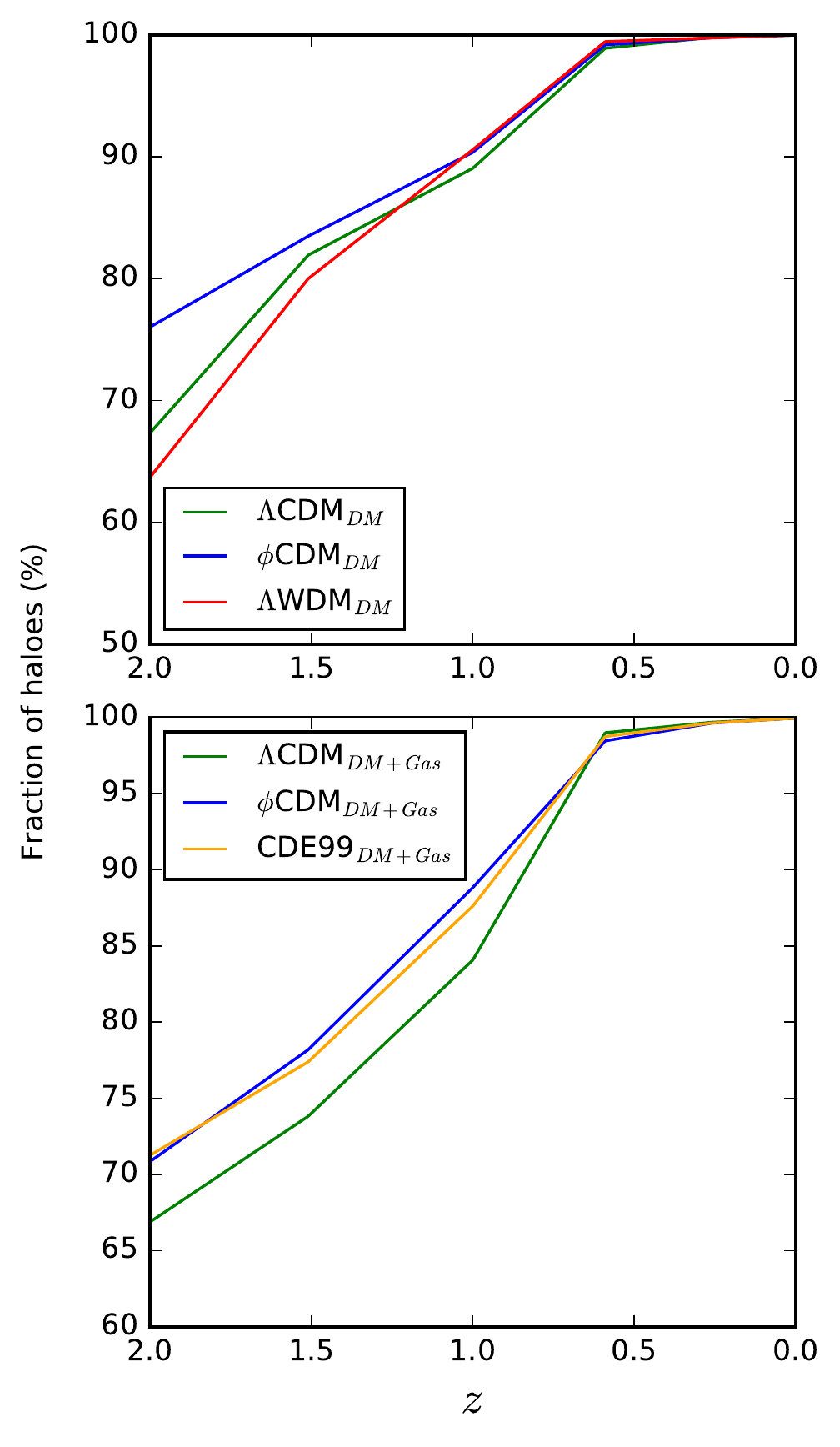}
\caption{The top and bottom panels are for DM and DM+Gas suites, respectively. Each line tracks the percentage of haloes residing within high density regions which are defined as clusters or filaments. \qcdmdm{} shows a significant amount of haloes migrating to $z$=0 from high density regions compared with \lcdmdm{} and \lwdmdm{} haloes.}\label{figmigtlly}
\end{figure}

Tracking the movement of haloes (excluding subhaloes) from snapshot to snapshot and recording in which LSS they reside at each time step allows the migration patterns of haloes to be measured and compared across cosmologies. We track the migration of $z$=0 haloes for DM (top row) and DM+Gas (bottom row) simulations from $z$=0 - 2.98 (or until formation time, to be elucidated below) in Figure \ref{figmigtlly}. Each line height (colored by cosmology) represents the percentage of total number of haloes within the particular redshift. We define the LSS type by fixing the simulation mask to be at $z$=0 and $\sigma_{s}$=2$\textit{h}^{-1}$Mpc for each respective cosmology. We then track haloes from low to high redshift by tracing the progenitors of parent haloes (using \textsc{treefrog} as elucidated in Section \ref{methhalos}) and tally in which LSS they reside at which redshift. We find a significant difference between the quintessence cosmologies \qcdmdm{}/\qcdmg{} where a higher quantity originate from filament and cluster regions as compared with the other cosmologies at $z_{mt}\geq$1.51. Summarizing the fractions of haloes residing within filaments and clusters, classed as high density regions, for DM and DM+Gas simulations seen in Figure \ref{figmigtlly} further highlights the difference between \qcdmdm{} and \lcdmdm{} especially. Given there are more haloes found within \qcdmdm{}/\qcdmg{} as compared with \lcdm{}, we test whether this higher fraction of haloes at high redshift is a consequence of a higher quantity of haloes. We sample the \qcdmdm{}/\qcdmg{} halo catalogs such that the mass distribution of haloes emulates that of \lcdm{}, but we find virtually identical fractions as in Figure \ref{figmigtlly}.

Given that haloes in the quintessence cosmology spend more time within high density regions, we measure the effects this may have on halo formation time. There are multiple definitions of halo formation time ($z_{ft}$). We define it as the time when half the total mass of the $z$=0 halo is acquired as per \citet{Sheth_04}, although other similar definitions have been used \citep[e.g][]{Li_08}. Using \textsc{treefrog} we identified progenitors by finding the candidate with the highest merit, as discussed in Section \ref{methalgnstat}, until the progenitor, which is at least $M_{z_0}/M_{z_{ft}}\geq 2$, is reached. 
The top (DM simulations) and bottom (DM+Gas simulations) panels of Figure \ref{figft} show the formation time $z$ of mass-averaged haloes (excluding subhaloes as they may gain and lose mass over time) where the mean and 1$\sigma$ is taken for each mass bin via bootstrap resampling. Clearly the trends for all DM and DM+Gas cosmologies show that higher mass haloes form later than low mass haloes, in agreement with the likes of \citet{Lacey_93,power_12,Elahi_18}. Noticeably the uncoupled quintessence cosmologies (represented in blue) seem to peak above the rest at $10^{13.5}$ - $10^{14.25}$ $\Msun{}$.

 
\section{Discussion}\label{discsec} 

\subsection{Spin Alignment}\label{disclsa} 
The spin-filament signal is a widely published alignment correlation which does not fully adhere to TTT predictions at low halo masses. We find within both DM and DM+Gas simulations (Figure \ref{figspinfildm} and \ref{figspinfildmg}) at all redshifts and smoothing scales, there is a mass dependence on the alignment strength. High mass haloes tend to be orthogonally aligned, thus more in tune with TTT predictions, whereas low mass haloes are closer to random alignment. \citet{Trowland_13, Wang_17} have shown, among many others, that low mass haloes have a parallel alignment (not seen within our signals as we don't have the mass resolution after halo particle quantity cut-off, see Section \ref{methhalos} for details). \citet{Codis_12} and \citet{Wang_17} conclude from simulations that parallel alignment of low mass haloes is due to their forming at earlier times than haloes at higher mass. In effect, as they migrate to filaments, already formed, they are less susceptible than unformed haloes to the tidal field, thus their spin alignment remains parallel. Wheres high mass haloes tend to undergo mergers within filaments, which orthogonally aligns their spins with respect to their residing filaments \citep{Welker_14}. Our haloes shown in Figure \ref{figft} for all cosmologies experience a mass dependency whereby, high mass haloes form later than low mass haloes, thus allowing them to be influenced by the filament's tidal field and which in turn strengthens their orthogonal alignment. This reaffirms the notion that formation time has a significant influence on spin-filament alignment \citep{Wang_17,Welker_14,Codis_12}. These figures show that alignments resulting from gravitational torques are to first order, not significantly affected by changes to dark sector physics. Although, haloes within \qcdmdm{} do form earlier than their counterparts within other cosmologies to be further discussed within Section \ref{discmm}.

Smoothing our simulations to various scales leads to different classifications of the cosmic web due to its hierarchical structure (noticeable in \citet{colless_01}). Within Section \ref{reslsa} and Figure \ref{figspinfildmg} we highlighted the smoothing scale-based trend whereby low redshift haloes are stronger in orthogonal alignment at 2$\textit{h}^{-1}$Mpc smoothing and gradually decreasing up to 5$\textit{h}^{-1}$Mpc, but at high redshift (from $z$=1.00) this trend reverses. We can conclude that haloes, being stronger aligned to smaller filaments at early times and larger filaments at late times, provide a strong indication of the growth of filaments (width-wise) over time, in agreement with \citet{Trowland_13}. Inspecting cluster haloes (plots were omitted for brevity), we find a distinct trend whereby $\sigma_{s}$=2$\textit{h}^{-1}$Mpc remains the dominant scale at which orthogonal alignment is strongest for all redshifts. We emphasize that non-standard cosmologies also mimic the smoothing scale trends highlighted within Figure \ref{figspinfildmg} which is a testament to the small effects non-standard dark sector physics have on halo spin-filament alignment and LSS evolution. Although, there are noticeable differences in the individual signals of non-standard cosmologies, as compared with \lcdm{}, which can be traced back to the differing dark sectors (as all simulations had the same density perturbation phases and due to the size of the simulation box, cosmic variance would be insignificant) but tracing back such fluctuations may not be worthwhile as they are unsystematic and inconsistent to be uniquely characterized.

High density regions such as clusters, filaments and sheets are regions where haloes are most susceptible to non-linear physics such as mergers and accretion. As described within Section \ref{reslsa}, we filter out haloes depending upon their proximity to voids and observe the spin alignment with the local tidal field. Figure \ref{figvsspinlssdm} alignments largely mimic those of high density regions, whereby at higher halo masses and redshift, alignments are strongly orthogonal, for all void region distance ranges. There is a significant difference between \qcdmdm{} and other cosmologies at $D_{v}$=2.4 $\textit{h}^{-1}$Mpc for $z$=0.59 shown in the bottom right panel, also within the middle panel, but differences are inconsistent and difficult to characterize. Within the DM+Gas simulations (Figure \ref{figvsspinlssdmg}) we identified \lcdmg{} as consistently stronger in orthogonal alignment compared with \qcdmg{}. They are maximally different at $\log(\Msun)\geq 13.5$ for $z$=1.00 at long distances. This could be a signature of the varied evacuation rate of particles out of void regions for \qcdmg{} (suggested by \citet{Adermann_17}) which lead to a disruption in the tidal field $\textbf{e}_3$, thus weakening the halo-LSS alignment.

Halo spin-spin alignments are also predicted by TTT to be parallel aligned at close enough distances, thus differing tidal fields could produce differing alignments between haloes. We use halo spin-spin as a probe of cosmology as signatures could arise from distinct spin-up from tidal fields of non-standard cosmologies. Figure \ref{fighhspin} shows some strong alignment between haloes at distances of $\log_{10}{[r/(\textit{h}^{-1}Mpc)]}<0$ for DM simulations (right column) and $\log_{10}{[r/(\textit{h}^{-1}Mpc)]}<-0.4$ for DM+Gas simulations (left column) at low redshift. As we proceed to higher redshift for DM simulations, the parallel alignment gradually fades, becoming completely random at $z$=1.00. We take these alignments (Figure \ref{fighhspin}) without filtering by LSS type, but if we filter out for filament haloes or haloes within void regions (plots are omitted for brevity) we see much weaker alignment signals, with little difference between individual cosmologies for DM or DM+Gas simulations. Since these alignments are largely found at small scales within FoF haloes \citep{Trowland_13}, we speculate that the clustering of small haloes (substructure) within high density regions begins at later times $z<1.00$, so it is only then that haloes are within range for tidal interactions to generate parallel spin alignments. This would explain the random alignment of haloes at high redshift.

\subsection{Movement and Migration}\label{discmm}
Strong parallel alignments of halo velocity-filament hint that haloes are streaming down filaments, in agreement with \citet{Trowland_13} findings. High mass haloes display stronger parallel alignment (Figure \ref{figvelfildmg}) which reaffirms the notion of these haloes forming within filaments. As such, strong velocity alignments (compared with low mass haloes) are an indication that haloes experience mergers and accretion along the direction of their respective filament axes. 

The alignment between halo spin and velocity (for haloes residing within filaments) displayed within Figure \ref{figspinveldmg} shows a considerably weaker alignment than expected, given the high strength of parallel alignment between velocity and filaments, and the orthogonal alignment between spin and filaments. We do not see the systematic shift towards orthogonal alignment at higher redshift which is generally seen for other filament halo alignments. Despite massive haloes becoming increasingly orthogonally aligned with decreasing redshift. This shows tentative evidence that regardless of filament axes, a haloes' spin will be aligned (orthogonally) with its velocity, purely based on the mass of the halo. This is reaffirmed by the cluster haloes in Figure \ref{figspinveldmg} having a mass dependency. Perhaps it is not only within filaments but all LSS that mergers and accretion realigns halo spins with their LSS axes. 

Once again we extend our search for cosmological signatures to the outskirts of voids, as shown in Figure \ref{figvsvellssdmg}, where haloes are less likely to be disturbed. Within the outskirts of voids, the velocity alignment to $e_{3}$ is no longer mass dependent; but is still somewhat parallel aligned. This is not unique to void region haloes but is common within all LSS, barring filaments. Nevertheless we find no unique cosmological signatures within velocity alignment analysis despite the differing LSS evolution with \qcdm{} concluded by \citet{Adermann_17,Watts_17}.

\begin{figure}
\centering
\includegraphics[width=0.35\textwidth]{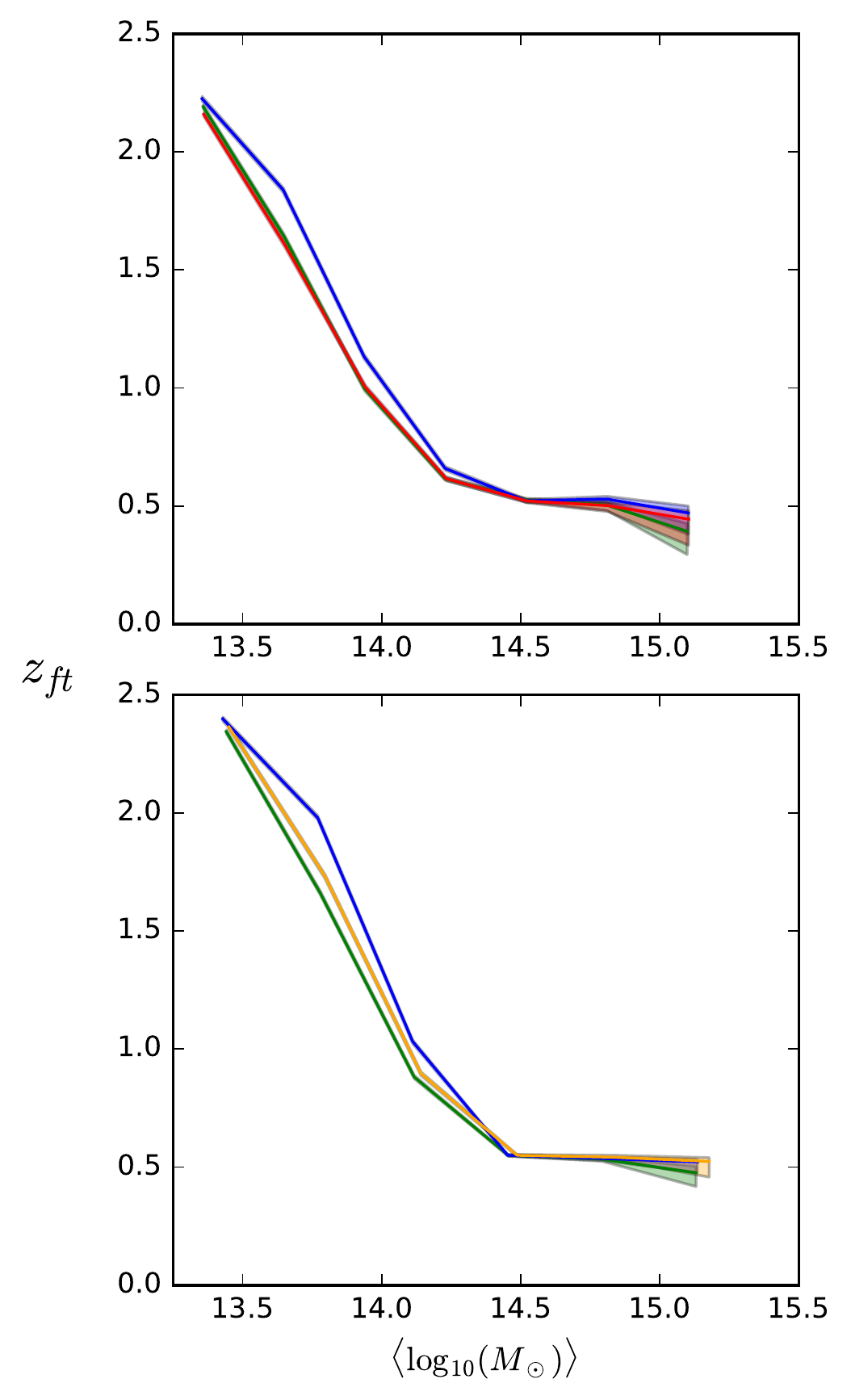}
\caption{The formation time of haloes, defined as the redshift at which haloes acquire half their $z$=0 mass. DM simulations (top panel) and DM+Gas simulations (bottom panel) show a mass dependence of low mass haloes forming earlier than high mass haloes with both \qcdmdm{} and \qcdmg{} haloes forming the earliest. Note: \cdeoodmg{} is omitted from this figure as earlier forming haloes are unfairly represented due to missing $z$=2.98 snapshot.}\label{figft}
\end{figure}

There is a clear formation time difference between \lcdm{} and \qcdm{} for both simulation suites (in Figure \ref{figft}). Moreover, the quintessence cosmology finds that a higher fraction of haloes to be within high density regions at high redshift (Figure \ref{figmigtlly}). These two discrepancies could be associated in the form of an assembly bias for quintessence haloes. This is speculative as although \citet{Watts_17} concluded higher cluster abundances within the DM simulation suite, it is unknown whether this higher clustering is significant enough to cause bias without further investigation. 

It is possible the earlier formation of haloes within \qcdm{} is due to an influx in mergers and accretion. On the basis of a higher merger rate \citep[which is also environmentally dependant][]{Jian_12} it is possible that cosmologies could be distinguished as mergers are thought to influence galaxy properties such as star formation rate, colour and morphology \citep{toomre_72}. 

The quintessence scalar field drives matter out of void regions into higher density regions at an increased rate. This extra force, with respect to \lcdm{}, would be expected to also accelerate the formation of haloes as seen in Figure \ref{figft}, produce more haloes which explains why a larger fraction of haloes migrate from relatively higher density regions as compared with \lcdm{}. Interestingly, with regards to the discrepent migration of haloes for \qcdmdm{}, the trend is not mimicked by \qcdmg{} which further begs the question, are these differences in a quintessence cosmology observable?

\section{Conclusion}\label{concsec} 
We have compared halo statistics over the redshift range $z$=0.00-2.98 between \lcdm{} cosmology and five non-standard cosmologies: two from DM only simulations, \lwdmdm{} and \qcdmdm{}, and three from hydrodynamical simulations, \qcdmg{}, \cdeodmg{} and \cdeoodmg{}. Specifically, the statistics we compared, in an effort to distinguish non-standard cosmologies, are halo spin/velocity alignment between LSS and between neighboring haloes, halo formation time and the spatial migration of haloes across various LSS. 

Firstly, we find that a higher percentage of $z$=0 \qcdmdm{} haloes originate from filament and cluster regions as compared with \lcdmdm{} and \lwdmdm{} by as much as 10 percent of the total population at high redshift. Migration differences of haloes between cosmologies would suggest a differing evolution history of LSS and/or of the rate of matter flow between LSS. By the same token we find that \qcdmdm{} haloes form earlier than their counterparts within the other DM cosmologies. These two findings may have some relation: perhaps haloes are spending more time within high density regions and are forming more rapidly within the \qcdmdm{} cosmology. Crucially \citet{Watts_17} compared the same DM simulation suite as used for this investigation and find there are clear distinctions between \qcdmdm{} and concordance. Specifically, they conclude that there are higher cluster abundances and lower void abundances within \qcdmdm{}, which are characteristically different from tweaking \lcdmdm{} cosmological parameters. With regards to the DM+Gas simulations analysed in this investigation, we find a weaker fraction discrepancy of \qcdmg{} halo migration at high redshifts, although \qcdmg{} haloes form significantly earlier than the fiducial cosmology similarly to \qcdmdm{}. It could be beneficial in future work to measure clustering and the merger rate of haloes for DM simulation suite in order to quantify any differences between cosmologies that might exist and determine whether there is an assembly bias in \qcdm{}.

In light of the first conclusion, we expected there would be some discrepancy between halo spin-filament alignment for \qcdmdm{} haloes. Despite TTT predictions of unformed haloes being susceptible and spun-up by the tidal field, the distinct formation times of \qcdmdm{} haloes and their evolutionary pathway, we still find that they are in full agreement with \lcdmdm{} halo alignments. We find for all cosmologies (DM and DM+Gas simulations) the alignment of spin with filament axes has a mass dependency. That is, high mass haloes have a stronger orthogonal alignment than low mass haloes. This suggests that low mass haloes, having formed earlier than high mass haloes, are less susceptible to the tidal field within filaments. High mass haloes having been found to form later, merge and accrete along the filament axes, flipping and/or strengthening their orthogonal alignment. High redshift alignments are systematically stronger (orthogonally) at low smoothing scale, whereas at low redshift they are stronger at high smoothing scale. Given that stronger orthogonal alignment of spin with a filament axis suggests a better defined filament, we conclude that filaments are growing, width-wise, over time. Finally, we inspect low density regions of our simulations as high density regions are highly non-linear. Halo-spin alignment within low density regions show little cosmological differences, albeit they are noisier due to a scarcity of haloes. In summary we do not find any cosmological signatures within spin-LSS alignments.

TTT predicts that haloes within proximity should hold some parallel alignment between their spins. We find these halo spin-spin alignments to be consistent across cosmologies as found within spin-filament alignments. We find no systematic differences between cosmologies, although there are noteworthy discrepancies between the halo spin-spin alignment distributions.

In future work, higher resolution simulations on the order of $1024^{3}$ particles could be beneficial in this case as cosmological signatures, being imprinted on halo spins, are a second-order effect. Thus a more in-depth investigation will be the nature of future work on seeking cosmological signatures of non-standrard cosmologies.

\section*{Acknowledgments}
The authors acknowledge the University of Sydney HPC services including the Artemis cluster, for providing computational resources that have contributed to the research results reported within this paper. We would also like to give special thanks to Peng Wang for his assistance with regards to the implementation of the Hessian matrix method.

\bibliographystyle{mnras}
\bibliography{biblio} 

\begin{thebibliography}{}
\makeatletter
\relax
\def\mn@urlcharsother{\let\do\@makeother \do\$\do\&\do\#\do\^\do\_\do\%\do\~}
\def\mn@doi{\begingroup\mn@urlcharsother \@ifnextchar [ {\mn@doi@}
  {\mn@doi@[]}}
\def\mn@doi@[#1]#2{\def\@tempa{#1}\ifx\@tempa\@empty \href
  {http://dx.doi.org/#2} {doi:#2}\else \href {http://dx.doi.org/#2} {#1}\fi
  \endgroup}
\def\mn@eprint#1#2{\mn@eprint@#1:#2::\@nil}
\def\mn@eprint@arXiv#1{\href {http://arxiv.org/abs/#1} {{\tt arXiv:#1}}}
\def\mn@eprint@dblp#1{\href {http://dblp.uni-trier.de/rec/bibtex/#1.xml}
  {dblp:#1}}
\def\mn@eprint@#1:#2:#3:#4\@nil{\def\@tempa {#1}\def\@tempb {#2}\def\@tempc
  {#3}\ifx \@tempc \@empty \let \@tempc \@tempb \let \@tempb \@tempa \fi \ifx
  \@tempb \@empty \def\@tempb {arXiv}\fi \@ifundefined
  {mn@eprint@\@tempb}{\@tempb:\@tempc}{\expandafter \expandafter \csname
  mn@eprint@\@tempb\endcsname \expandafter{\@tempc}}}

\bibitem[\protect\citeauthoryear{{Abazajian} et~al.,}{{Abazajian}
  et~al.}{2009}]{Abazajian_09}
{Abazajian} K.~N.,  et~al., 2009, \mn@doi [\apjs]
  {10.1088/0067-0049/182/2/543}, \href
  {http://adsabs.harvard.edu/abs/2009ApJS..182..543A} {182, 543}

\bibitem[\protect\citeauthoryear{{Adermann}, {Elahi}, {Lewis}  \&
  {Power}}{{Adermann} et~al.}{2017}]{Adermann_17}
{Adermann} E.,  {Elahi} P.~J.,  {Lewis} G.~F.,   {Power} C.,  2017, \mn@doi
  [\mnras] {10.1093/mnras/stx657}, \href
  {http://adsabs.harvard.edu/abs/2017MNRAS.468.3381A} {468, 3381}

\bibitem[\protect\citeauthoryear{{Adermann}, {Elahi}, {Lewis}  \&
  {Power}}{{Adermann} et~al.}{2018}]{Adermann_18}
{Adermann} E.,  {Elahi} P.~J.,  {Lewis} G.~F.,   {Power} C.,  2018, \mn@doi
  [\mnras] {10.1093/mnras/sty1824}, \href
  {https://ui.adsabs.harvard.edu/abs/2018MNRAS.479.4861A} {479, 4861}

\bibitem[\protect\citeauthoryear{{Aragon-Calvo} \& {Yang}}{{Aragon-Calvo} \&
  {Yang}}{2014}]{Calvo_14}
{Aragon-Calvo} M.~A.,  {Yang} L.~F.,  2014, \mn@doi [\mnras]
  {10.1093/mnrasl/slu009}, \href
  {http://adsabs.harvard.edu/abs/2014MNRAS.440L..46A} {440, L46}

\bibitem[\protect\citeauthoryear{{Arag{\'o}n-Calvo}, {Jones}, {van de Weygaert}
   \& {van der Hulst}}{{Arag{\'o}n-Calvo} et~al.}{2007}]{Calvo_07}
{Arag{\'o}n-Calvo} M.~A.,  {Jones} B.~J.~T.,  {van de Weygaert} R.,   {van der
  Hulst} J.~M.,  2007, \mn@doi [\aap] {10.1051/0004-6361:20077880}, \href
  {http://adsabs.harvard.edu/abs/2007A%26A...474..315A} {474, 315}

\bibitem[\protect\citeauthoryear{{Bailin} \& {Steinmetz}}{{Bailin} \&
  {Steinmetz}}{2005}]{Bailin_05}
{Bailin} J.,  {Steinmetz} M.,  2005, \mn@doi [\apj] {10.1086/430397}, \href
  {http://adsabs.harvard.edu/abs/2005ApJ...627..647B} {627, 647}

\bibitem[\protect\citeauthoryear{{Baldi}}{{Baldi}}{2012}]{Baldi_12}
{Baldi} M.,  2012, \mn@doi [\mnras] {10.1111/j.1365-2966.2012.20675.x}, \href
  {http://adsabs.harvard.edu/abs/2012MNRAS.422.1028B} {422, 1028}

\bibitem[\protect\citeauthoryear{{Baldi}, {Pettorino}, {Robbers}  \&
  {Springel}}{{Baldi} et~al.}{2010}]{Baldi_10}
{Baldi} M.,  {Pettorino} V.,  {Robbers} G.,   {Springel} V.,  2010, \mn@doi
  [\mnras] {10.1111/j.1365-2966.2009.15987.x}, \href
  {http://adsabs.harvard.edu/abs/2010MNRAS.403.1684B} {403, 1684}

\bibitem[\protect\citeauthoryear{{Bennett} et~al.,}{{Bennett}
  et~al.}{2013}]{Bennett_13}
{Bennett} C.~L.,  et~al., 2013, \mn@doi [\apjs] {10.1088/0067-0049/208/2/20},
  \href {http://adsabs.harvard.edu/abs/2013ApJS..208...20B} {208, 20}

\bibitem[\protect\citeauthoryear{{Bertone}, {Hooper}  \& {Silk}}{{Bertone}
  et~al.}{2005}]{Bertone_05}
{Bertone} G.,  {Hooper} D.,   {Silk} J.,  2005, \mn@doi [\physrep]
  {10.1016/j.physrep.2004.08.031}, \href
  {http://adsabs.harvard.edu/abs/2005PhR...405..279B} {405, 279}

\bibitem[\protect\citeauthoryear{{Beutler} et~al.,}{{Beutler}
  et~al.}{2011}]{Beutler_11}
{Beutler} F.,  et~al., 2011, \mn@doi [\mnras]
  {10.1111/j.1365-2966.2011.19250.x}, \href
  {http://adsabs.harvard.edu/abs/2011MNRAS.416.3017B} {416, 3017}

\bibitem[\protect\citeauthoryear{{Bode}, {Ostriker}  \& {Turok}}{{Bode}
  et~al.}{2001}]{Bode_01}
{Bode} P.,  {Ostriker} J.~P.,   {Turok} N.,  2001, \mn@doi [\apj]
  {10.1086/321541}, \href {http://adsabs.harvard.edu/abs/2001ApJ...556...93B}
  {556, 93}

\bibitem[\protect\citeauthoryear{{Bond}, {Kofman}  \& {Pogosyan}}{{Bond}
  et~al.}{1996}]{Bond_96}
{Bond} J.~R.,  {Kofman} L.,   {Pogosyan} D.,  1996, \mn@doi [\nat]
  {10.1038/380603a0}, \href {http://adsabs.harvard.edu/abs/1996Natur.380..603B}
  {380, 603}

\bibitem[\protect\citeauthoryear{{Bull} et~al.,}{{Bull} et~al.}{2016}]{Bull_16}
{Bull} P.,  et~al., 2016, \mn@doi [Physics of the Dark Universe]
  {10.1016/j.dark.2016.02.001}, \href
  {http://adsabs.harvard.edu/abs/2016PDU....12...56B} {12, 56}

\bibitem[\protect\citeauthoryear{{Bullock}, {Kravtsov}  \&
  {Weinberg}}{{Bullock} et~al.}{2000}]{Bullock_00}
{Bullock} J.~S.,  {Kravtsov} A.~V.,   {Weinberg} D.~H.,  2000, \mn@doi [\apj]
  {10.1086/309279}, \href {http://adsabs.harvard.edu/abs/2000ApJ...539..517B}
  {539, 517}

\bibitem[\protect\citeauthoryear{{Carlesi}, {Knebe}, {Lewis}, {Wales}  \&
  {Yepes}}{{Carlesi} et~al.}{2014a}]{Carlesi_14a}
{Carlesi} E.,  {Knebe} A.,  {Lewis} G.~F.,  {Wales} S.,   {Yepes} G.,  2014a,
  \mn@doi [\mnras] {10.1093/mnras/stu150}, \href
  {http://adsabs.harvard.edu/abs/2014MNRAS.439.2943C} {439, 2943}

\bibitem[\protect\citeauthoryear{{Carlesi}, {Knebe}, {Lewis}  \&
  {Yepes}}{{Carlesi} et~al.}{2014b}]{Carlesi_14b}
{Carlesi} E.,  {Knebe} A.,  {Lewis} G.~F.,   {Yepes} G.,  2014b, \mn@doi
  [\mnras] {10.1093/mnras/stu151}, \href
  {http://adsabs.harvard.edu/abs/2014MNRAS.439.2958C} {439, 2958}

\bibitem[\protect\citeauthoryear{{Cautun} \& {van de Weygaert}}{{Cautun} \&
  {van de Weygaert}}{2011}]{Cautun_11}
{Cautun} M.~C.,  {van de Weygaert} R.,  2011, {The DTFE public software: The
  Delaunay Tessellation Field Estimator code}, Astrophysics Source Code Library
  (\mn@eprint {arXiv} {1105.0370})

\bibitem[\protect\citeauthoryear{{Chuang} et~al.,}{{Chuang}
  et~al.}{2016}]{Chuang_16}
{Chuang} C.-H.,  et~al., 2016, \mn@doi [\mnras] {10.1093/mnras/stw1535}, \href
  {https://ui.adsabs.harvard.edu/\#abs/2016MNRAS.461.3781C} {461, 3781}

\bibitem[\protect\citeauthoryear{{Codis}, {Pichon}, {Devriendt}, {Slyz},
  {Pogosyan}, {Dubois}  \& {Sousbie}}{{Codis} et~al.}{2012}]{Codis_12}
{Codis} S.,  {Pichon} C.,  {Devriendt} J.,  {Slyz} A.,  {Pogosyan} D.,
  {Dubois} Y.,   {Sousbie} T.,  2012, \mn@doi [\mnras]
  {10.1111/j.1365-2966.2012.21636.x}, \href
  {http://adsabs.harvard.edu/abs/2012MNRAS.427.3320C} {427, 3320}

\bibitem[\protect\citeauthoryear{{Colless} et~al.,}{{Colless}
  et~al.}{2001}]{colless_01}
{Colless} M.,  et~al., 2001, \mn@doi [\mnras]
  {10.1046/j.1365-8711.2001.04902.x}, \href
  {http://adsabs.harvard.edu/abs/2001MNRAS.328.1039C} {328, 1039}

\bibitem[\protect\citeauthoryear{{Doran}, {Karwan}  \& {Wetterich}}{{Doran}
  et~al.}{2005}]{Doran_05}
{Doran} M.,  {Karwan} K.,   {Wetterich} C.,  2005, \mn@doi [\jcap]
  {10.1088/1475-7516/2005/11/007}, \href
  {http://adsabs.harvard.edu/abs/2005JCAP...11..007D} {11, 007}

\bibitem[\protect\citeauthoryear{{Dubois} et~al.,}{{Dubois}
  et~al.}{2014}]{Dubois_14}
{Dubois} Y.,  et~al., 2014, \mn@doi [\mnras] {10.1093/mnras/stu1227}, \href
  {http://adsabs.harvard.edu/abs/2014MNRAS.444.1453D} {444, 1453}

\bibitem[\protect\citeauthoryear{{Elahi}, {Thacker}  \& {Widrow}}{{Elahi}
  et~al.}{2011}]{Elahi_11}
{Elahi} P.~J.,  {Thacker} R.~J.,   {Widrow} L.~M.,  2011, \mn@doi [\mnras]
  {10.1111/j.1365-2966.2011.19485.x}, \href
  {http://adsabs.harvard.edu/abs/2011MNRAS.418..320E} {418, 320}

\bibitem[\protect\citeauthoryear{{Elahi}, {Lewis}, {Power}, {Carlesi}  \&
  {Knebe}}{{Elahi} et~al.}{2015}]{Elahi_15}
{Elahi} P.~J.,  {Lewis} G.~F.,  {Power} C.,  {Carlesi} E.,   {Knebe} A.,  2015,
  \mn@doi [\mnras] {10.1093/mnras/stv1370}, \href
  {http://adsabs.harvard.edu/abs/2015MNRAS.452.1341E} {452, 1341}

\bibitem[\protect\citeauthoryear{{Elahi}, {Welker}, {Power}, {Lagos},
  {Robotham}, {Ca{\~n}as}  \& {Poulton}}{{Elahi} et~al.}{2018}]{Elahi_18}
{Elahi} P.~J.,  {Welker} C.,  {Power} C.,  {Lagos} C.~d.~P.,  {Robotham}
  A.~S.~G.,  {Ca{\~n}as} R.,   {Poulton} R.,  2018, \mn@doi [\mnras]
  {10.1093/mnras/sty061}, \href
  {http://adsabs.harvard.edu/abs/2018MNRAS.475.5338E} {475, 5338}

\bibitem[\protect\citeauthoryear{{Elahi}, {Ca{\~n}as}, {Tobar}, {Willis},
  {Lagos}, {Power}  \& {Robotham}}{{Elahi} et~al.}{2019a}]{Elahi_19a}
{Elahi} P.~J.,  {Ca{\~n}as} R.,  {Tobar} R.~J.,  {Willis} J.~S.,  {Lagos} C.
  d.~P.,  {Power} C.,   {Robotham} A. S.~G.,  2019a, arXiv e-prints, \href
  {https://ui.adsabs.harvard.edu/\#abs/2019arXiv190201010E} {p.
  arXiv:1902.01010}

\bibitem[\protect\citeauthoryear{{Elahi}, {Poulton}, {Tobar}, {Lagos}, {Power}
  \& {Robotham}}{{Elahi} et~al.}{2019b}]{Elahi_19b}
{Elahi} P.~J.,  {Poulton} R. J.~J.,  {Tobar} R.~J.,  {Lagos} C. d.~P.,  {Power}
  C.,   {Robotham} A. S.~G.,  2019b, arXiv e-prints, \href
  {https://ui.adsabs.harvard.edu/\#abs/2019arXiv190201527E} {p.
  arXiv:1902.01527}

\bibitem[\protect\citeauthoryear{{Faltenbacher}, {Gottl{\"o}ber}, {Kerscher}
  \& {M{\"u}ller}}{{Faltenbacher} et~al.}{2002}]{Faltenbacher_02}
{Faltenbacher} A.,  {Gottl{\"o}ber} S.,  {Kerscher} M.,   {M{\"u}ller} V.,
  2002, \mn@doi [\aap] {10.1051/0004-6361:20021263}, \href
  {http://adsabs.harvard.edu/abs/2002A%26A...395....1F} {395, 1}

\bibitem[\protect\citeauthoryear{{Ganeshaiah Veena}, {Cautun}, {van de
  Weygaert}, {Tempel}, {Jones}, {Rieder}  \& {Frenk}}{{Ganeshaiah Veena}
  et~al.}{2018}]{Veena_18}
{Ganeshaiah Veena} P.,  {Cautun} M.,  {van de Weygaert} R.,  {Tempel} E.,
  {Jones} B.~J.~T.,  {Rieder} S.,   {Frenk} C.~S.,  2018, preprint, \href
  {http://adsabs.harvard.edu/abs/2018arXiv180500033G} {} (\mn@eprint {arXiv}
  {1805.00033})

\bibitem[\protect\citeauthoryear{{Hahn}, {Porciani}, {Carollo}  \&
  {Dekel}}{{Hahn} et~al.}{2007}]{Hahn_07}
{Hahn} O.,  {Porciani} C.,  {Carollo} C.~M.,   {Dekel} A.,  2007, \mn@doi
  [\mnras] {10.1111/j.1365-2966.2006.11318.x}, \href
  {http://adsabs.harvard.edu/abs/2007MNRAS.375..489H} {375, 489}

\bibitem[\protect\citeauthoryear{{Hoyle}, {Burgers}, {van de Hulst}  \&
  eds}{{Hoyle} et~al.}{1949}]{hoyle_1949}
{Hoyle} F.,  {Burgers} J.,  {van de Hulst} H.,   eds 1949, Central Air
  Documents Office, Dayton, p.~195

\bibitem[\protect\citeauthoryear{{Ibata} et~al.,}{{Ibata}
  et~al.}{2013}]{Ibata_13}
{Ibata} R.~A.,  et~al., 2013, \mn@doi [\nat] {10.1038/nature11717}, \href
  {https://ui.adsabs.harvard.edu/abs/2013Natur.493...62I} {493, 62}

\bibitem[\protect\citeauthoryear{{Jian}, {Lin}  \& {Chiueh}}{{Jian}
  et~al.}{2012}]{Jian_12}
{Jian} H.-Y.,  {Lin} L.,   {Chiueh} T.,  2012, \mn@doi [\apj]
  {10.1088/0004-637X/754/1/26}, \href
  {https://ui.adsabs.harvard.edu/abs/2012ApJ...754...26J} {754, 26}

\bibitem[\protect\citeauthoryear{{Jones}, {van de Weygaert}  \&
  {Arag{\'o}n-Calvo}}{{Jones} et~al.}{2010}]{Jones_10}
{Jones} B.~J.~T.,  {van de Weygaert} R.,   {Arag{\'o}n-Calvo} M.~A.,  2010,
  \mn@doi [\mnras] {10.1111/j.1365-2966.2010.17202.x}, \href
  {http://adsabs.harvard.edu/abs/2010MNRAS.408..897J} {408, 897}

\bibitem[\protect\citeauthoryear{{Joyce}, {Jain}, {Khoury}  \&
  {Trodden}}{{Joyce} et~al.}{2015}]{Joyce_15}
{Joyce} A.,  {Jain} B.,  {Khoury} J.,   {Trodden} M.,  2015, \mn@doi [\physrep]
  {10.1016/j.physrep.2014.12.002}, \href
  {http://adsabs.harvard.edu/abs/2015PhR...568....1J} {568, 1}

\bibitem[\protect\citeauthoryear{{Kang} \& {Wang}}{{Kang} \&
  {Wang}}{2015}]{Kang_15}
{Kang} X.,  {Wang} P.,  2015, \mn@doi [\apj] {10.1088/0004-637X/813/1/6}, \href
  {http://adsabs.harvard.edu/abs/2015ApJ...813....6K} {813, 6}

\bibitem[\protect\citeauthoryear{{Kilbinger} et~al.,}{{Kilbinger}
  et~al.}{2013}]{Kilbinger_13}
{Kilbinger} M.,  et~al., 2013, \mn@doi [\mnras] {10.1093/mnras/stt041}, \href
  {http://adsabs.harvard.edu/abs/2013MNRAS.430.2200K} {430, 2200}

\bibitem[\protect\citeauthoryear{{Klypin}, {Gottl{\"o}ber}, {Kravtsov}  \&
  {Khokhlov}}{{Klypin} et~al.}{1999}]{Klypin_99}
{Klypin} A.,  {Gottl{\"o}ber} S.,  {Kravtsov} A.~V.,   {Khokhlov} A.~M.,  1999,
  \mn@doi [\apj] {10.1086/307122}, \href
  {http://adsabs.harvard.edu/abs/1999ApJ...516..530K} {516, 530}

\bibitem[\protect\citeauthoryear{{Lacey} \& {Cole}}{{Lacey} \&
  {Cole}}{1993}]{Lacey_93}
{Lacey} C.,  {Cole} S.,  1993, \mn@doi [\mnras] {10.1093/mnras/262.3.627},
  \href {http://adsabs.harvard.edu/abs/1993MNRAS.262..627L} {262, 627}

\bibitem[\protect\citeauthoryear{{Lee}}{{Lee}}{2011}]{Lee_11}
{Lee} J.,  2011, \mn@doi [\apj] {10.1088/0004-637X/732/2/99}, \href
  {http://adsabs.harvard.edu/abs/2011ApJ...732...99L} {732, 99}

\bibitem[\protect\citeauthoryear{{Lee} \& {Erdogdu}}{{Lee} \&
  {Erdogdu}}{2007}]{Lee_Erdogdu_07}
{Lee} J.,  {Erdogdu} P.,  2007, \mn@doi [\apj] {10.1086/523351}, \href
  {http://adsabs.harvard.edu/abs/2007ApJ...671.1248L} {671, 1248}

\bibitem[\protect\citeauthoryear{{Lee} \& {Pen}}{{Lee} \&
  {Pen}}{2000}]{Lee_pen_00}
{Lee} J.,  {Pen} U.-L.,  2000, \mn@doi [\apjl] {10.1086/312556}, \href
  {http://adsabs.harvard.edu/abs/2000ApJ...532L...5L} {532, L5}

\bibitem[\protect\citeauthoryear{{Li}, {Mo}  \& {Gao}}{{Li}
  et~al.}{2008}]{Li_08}
{Li} Y.,  {Mo} H.~J.,   {Gao} L.,  2008, \mn@doi [\mnras]
  {10.1111/j.1365-2966.2008.13667.x}, \href
  {https://ui.adsabs.harvard.edu/abs/2008MNRAS.389.1419L} {389, 1419}

\bibitem[\protect\citeauthoryear{{Libeskind}, {Hoffman}, {Knebe}, {Steinmetz},
  {Gottl{\"o}ber}, {Metuki}  \& {Yepes}}{{Libeskind}
  et~al.}{2012}]{Libeskind_12}
{Libeskind} N.~I.,  {Hoffman} Y.,  {Knebe} A.,  {Steinmetz} M.,
  {Gottl{\"o}ber} S.,  {Metuki} O.,   {Yepes} G.,  2012, \mn@doi [\mnras]
  {10.1111/j.1745-3933.2012.01222.x}, \href
  {http://adsabs.harvard.edu/abs/2012MNRAS.421L.137L} {421, L137}

\bibitem[\protect\citeauthoryear{{Moore}, {Ghigna}, {Governato}, {Lake},
  {Quinn}, {Stadel}  \& {Tozzi}}{{Moore} et~al.}{1999}]{Moore_99}
{Moore} B.,  {Ghigna} S.,  {Governato} F.,  {Lake} G.,  {Quinn} T.,  {Stadel}
  J.,   {Tozzi} P.,  1999, \mn@doi [\apjl] {10.1086/312287}, \href
  {http://adsabs.harvard.edu/abs/1999ApJ...524L..19M} {524, L19}

\bibitem[\protect\citeauthoryear{{Pawlowski}, {Pflamm-Altenburg}  \&
  {Kroupa}}{{Pawlowski} et~al.}{2012}]{Pawloski_12}
{Pawlowski} M.~S.,  {Pflamm-Altenburg} J.,   {Kroupa} P.,  2012, \mn@doi
  [\mnras] {10.1111/j.1365-2966.2012.20937.x}, \href
  {http://adsabs.harvard.edu/abs/2012MNRAS.423.1109P} {423, 1109}

\bibitem[\protect\citeauthoryear{{Peebles}}{{Peebles}}{1969}]{Peebles_69}
{Peebles} P.~J.~E.,  1969, \mn@doi [\apj] {10.1086/149876}, \href
  {http://adsabs.harvard.edu/abs/1969ApJ...155..393P} {155, 393}

\bibitem[\protect\citeauthoryear{{Pen}, {Lee}  \& {Seljak}}{{Pen}
  et~al.}{2000}]{Pen_00}
{Pen} U.-L.,  {Lee} J.,   {Seljak} U.,  2000, \mn@doi [\apjl] {10.1086/317273},
  \href {http://adsabs.harvard.edu/abs/2000ApJ...543L.107P} {543, L107}

\bibitem[\protect\citeauthoryear{{Petraki} \& {Volkas}}{{Petraki} \&
  {Volkas}}{2013}]{Petraki_13}
{Petraki} K.,  {Volkas} R.~R.,  2013, \mn@doi [International Journal of Modern
  Physics A] {10.1142/S0217751X13300287}, \href
  {http://adsabs.harvard.edu/abs/2013IJMPA..2830028P} {28, 1330028}

\bibitem[\protect\citeauthoryear{{Pettorino}, {Amendola}, {Baccigalupi}  \&
  {Quercellini}}{{Pettorino} et~al.}{2012}]{Pettorino_12}
{Pettorino} V.,  {Amendola} L.,  {Baccigalupi} C.,   {Quercellini} C.,  2012,
  \mn@doi [\prd] {10.1103/PhysRevD.86.103507}, \href
  {http://adsabs.harvard.edu/abs/2012PhRvD..86j3507P} {86, 103507}

\bibitem[\protect\citeauthoryear{{Pichon}, {Codis}, {Pogosyan}, {Dubois},
  {Desjacques}  \& {Devriendt}}{{Pichon} et~al.}{2016}]{Pichon_16}
{Pichon} C.,  {Codis} S.,  {Pogosyan} D.,  {Dubois} Y.,  {Desjacques} V.,
  {Devriendt} J.,  2016, in {van de Weygaert} R.,  {Shandarin} S.,  {Saar} E.,
   {Einasto} J.,  eds,  IAU Symposium Vol. 308, The Zeldovich Universe: Genesis
  and Growth of the Cosmic Web. pp 421--432 (\mn@eprint {arXiv} {1409.2608}),
  \mn@doi{10.1017/S1743921316010309}

\bibitem[\protect\citeauthoryear{{Planck Collaboration} et~al.,}{{Planck
  Collaboration} et~al.}{2014}]{Plank_14b}
{Planck Collaboration} et~al., 2014, \mn@doi [\aap]
  {10.1051/0004-6361/201321534}, \href
  {http://adsabs.harvard.edu/abs/2014A%26A...571A..23P} {571, A23}

\bibitem[\protect\citeauthoryear{{Planck Collaboration} et~al.,}{{Planck
  Collaboration} et~al.}{2016}]{Plank_16}
{Planck Collaboration} et~al., 2016, \mn@doi [\aap]
  {10.1051/0004-6361/201526681}, \href
  {http://adsabs.harvard.edu/abs/2016A%26A...594A..16P} {594, A16}

\bibitem[\protect\citeauthoryear{{Power}, {Knebe}  \& {Knollmann}}{{Power}
  et~al.}{2012}]{power_12}
{Power} C.,  {Knebe} A.,   {Knollmann} S.~R.,  2012, \mn@doi [\mnras]
  {10.1111/j.1365-2966.2011.19820.x}, \href
  {https://ui.adsabs.harvard.edu/abs/2012MNRAS.419.1576P} {419, 1576}

\bibitem[\protect\citeauthoryear{{Ratra} \& {Peebles}}{{Ratra} \&
  {Peebles}}{1988}]{Ratra_88}
{Ratra} B.,  {Peebles} P.~J.~E.,  1988, \mn@doi [\prd]
  {10.1103/PhysRevD.37.3406}, \href
  {http://adsabs.harvard.edu/abs/1988PhRvD..37.3406R} {37, 3406}

\bibitem[\protect\citeauthoryear{{Schaap} \& {van de Weygaert}}{{Schaap} \&
  {van de Weygaert}}{2000}]{Shaap_00}
{Schaap} W.~E.,  {van de Weygaert} R.,  2000, \aap, \href
  {http://adsabs.harvard.edu/abs/2000A%26A...363L..29S} {363, L29}

\bibitem[\protect\citeauthoryear{{Sheth} \& {Tormen}}{{Sheth} \&
  {Tormen}}{2004}]{Sheth_04}
{Sheth} R.~K.,  {Tormen} G.,  2004, \mn@doi [\mnras]
  {10.1111/j.1365-2966.2004.07622.x}, \href
  {http://adsabs.harvard.edu/abs/2004MNRAS.349.1464S} {349, 1464}

\bibitem[\protect\citeauthoryear{{Slosar} et~al.,}{{Slosar}
  et~al.}{2009}]{Slosar_09}
{Slosar} A.,  et~al., 2009, \mn@doi [\mnras]
  {10.1111/j.1365-2966.2008.14127.x}, \href
  {http://adsabs.harvard.edu/abs/2009MNRAS.392.1225S} {392, 1225}

\bibitem[\protect\citeauthoryear{{Smith} \& {Markovic}}{{Smith} \&
  {Markovic}}{2011}]{Smith_11}
{Smith} R.~E.,  {Markovic} K.,  2011, \mn@doi [\prd]
  {10.1103/PhysRevD.84.063507}, \href
  {http://adsabs.harvard.edu/abs/2011PhRvD..84f3507S} {84, 063507}

\bibitem[\protect\citeauthoryear{{Springel}}{{Springel}}{2005}]{Springel_05}
{Springel} V.,  2005, \mn@doi [\mnras] {10.1111/j.1365-2966.2005.09655.x},
  \href {http://adsabs.harvard.edu/abs/2005MNRAS.364.1105S} {364, 1105}

\bibitem[\protect\citeauthoryear{{Suzuki} et~al.,}{{Suzuki}
  et~al.}{2012}]{Suzuki_12}
{Suzuki} N.,  et~al., 2012, \mn@doi [\apj] {10.1088/0004-637X/746/1/85}, \href
  {http://adsabs.harvard.edu/abs/2012ApJ...746...85S} {746, 85}

\bibitem[\protect\citeauthoryear{{Toomre} \& {Toomre}}{{Toomre} \&
  {Toomre}}{1972}]{toomre_72}
{Toomre} A.,  {Toomre} J.,  1972, \mn@doi [\apj] {10.1086/151823}, \href
  {https://ui.adsabs.harvard.edu/abs/1972ApJ...178..623T} {178, 623}

\bibitem[\protect\citeauthoryear{{Trowland}, {Lewis}  \&
  {Bland-Hawthorn}}{{Trowland} et~al.}{2013}]{Trowland_13}
{Trowland} H.~E.,  {Lewis} G.~F.,   {Bland-Hawthorn} J.,  2013, \mn@doi [\apj]
  {10.1088/0004-637X/762/2/72}, \href
  {http://adsabs.harvard.edu/abs/2013ApJ...762...72T} {762, 72}

\bibitem[\protect\citeauthoryear{{Tsujikawa}}{{Tsujikawa}}{2013}]{Tsujikawa_13}
{Tsujikawa} S.,  2013, \mn@doi [Classical and Quantum Gravity]
  {10.1088/0264-9381/30/21/214003}, \href
  {http://adsabs.harvard.edu/abs/2013CQGra..30u4003T} {30, 214003}

\bibitem[\protect\citeauthoryear{{Viel}, {Becker}, {Bolton}  \&
  {Haehnelt}}{{Viel} et~al.}{2013}]{Viel_13}
{Viel} M.,  {Becker} G.~D.,  {Bolton} J.~S.,   {Haehnelt} M.~G.,  2013, \mn@doi
  [\prd] {10.1103/PhysRevD.88.043502}, \href
  {http://adsabs.harvard.edu/abs/2013PhRvD..88d3502V} {88, 043502}

\bibitem[\protect\citeauthoryear{{Wang} \& {Kang}}{{Wang} \&
  {Kang}}{2017}]{Wang_17}
{Wang} P.,  {Kang} X.,  2017, \mn@doi [\mnras] {10.1093/mnrasl/slx038}, \href
  {http://adsabs.harvard.edu/abs/2017MNRAS.468L.123W} {468, L123}

\bibitem[\protect\citeauthoryear{{Watts}, {Elahi}, {Lewis}  \& {Power}}{{Watts}
  et~al.}{2017}]{Watts_17}
{Watts} A.~L.,  {Elahi} P.~J.,  {Lewis} G.~F.,   {Power} C.,  2017, \mn@doi
  [\mnras] {10.1093/mnras/stx375}, \href
  {http://adsabs.harvard.edu/abs/2017MNRAS.468...59W} {468, 59}

\bibitem[\protect\citeauthoryear{{Welker}, {Devriendt}, {Dubois}, {Pichon}  \&
  {Peirani}}{{Welker} et~al.}{2014}]{Welker_14}
{Welker} C.,  {Devriendt} J.,  {Dubois} Y.,  {Pichon} C.,   {Peirani} S.,
  2014, \mn@doi [\mnras] {10.1093/mnrasl/slu106}, \href
  {http://adsabs.harvard.edu/abs/2014MNRAS.445L..46W} {445, L46}

\bibitem[\protect\citeauthoryear{{Wetzel}, {Hopkins}, {Kim},
  {Faucher-Gigu{\`e}re}, {Kere{\v s}}  \& {Quataert}}{{Wetzel}
  et~al.}{2016}]{Wetzel_16}
{Wetzel} A.~R.,  {Hopkins} P.~F.,  {Kim} J.-h.,  {Faucher-Gigu{\`e}re} C.-A.,
  {Kere{\v s}} D.,   {Quataert} E.,  2016, \mn@doi [\apjl]
  {10.3847/2041-8205/827/2/L23}, \href
  {http://adsabs.harvard.edu/abs/2016ApJ...827L..23W} {827, L23}

\bibitem[\protect\citeauthoryear{{White}, {Frenk}  \& {Davis}}{{White}
  et~al.}{1983}]{White_83}
{White} S.~D.~M.,  {Frenk} C.~S.,   {Davis} M.,  1983, \mn@doi [\apjl]
  {10.1086/184139}, \href
  {https://ui.adsabs.harvard.edu/abs/1983ApJ...274L...1W} {274, L1}

\bibitem[\protect\citeauthoryear{{Zel'dovich}}{{Zel'dovich}}{1970}]{Zeldovich_70}
{Zel'dovich} Y.~B.,  1970, \aap, \href
  {http://adsabs.harvard.edu/abs/1970A%26A.....5...84Z} {5, 84}

\bibitem[\protect\citeauthoryear{{Zhang}, {Yang}, {Faltenbacher}, {Springel},
  {Lin}  \& {Wang}}{{Zhang} et~al.}{2009}]{Zhang_09}
{Zhang} Y.,  {Yang} X.,  {Faltenbacher} A.,  {Springel} V.,  {Lin} W.,   {Wang}
  H.,  2009, \mn@doi [\apj] {10.1088/0004-637X/706/1/747}, \href
  {http://adsabs.harvard.edu/abs/2009ApJ...706..747Z} {706, 747}

\bibitem[\protect\citeauthoryear{{van de Weygaert} \& {Schaap}}{{van de
  Weygaert} \& {Schaap}}{2009}]{Weygaert_09}
{van de Weygaert} R.,  {Schaap} W.,  2009, in {Mart{\'{\i}}nez} V.~J.,  {Saar}
  E.,  {Mart{\'{\i}}nez-Gonz{\'a}lez} E.,   {Pons-Border{\'{\i}}a} M.-J.,  eds,
   Lecture Notes in Physics, Berlin Springer Verlag Vol. 665, Data Analysis in
  Cosmology. pp 291--413 (\mn@eprint {arXiv} {0708.1441}),
  \mn@doi{10.1007/978-3-540-44767-2_11}

\makeatother
\end{thebibliography}

\bsp	
\label{lastpage}
\end{document}